\documentclass[a4paper,11pt]{article}
\pdfoutput=1
\usepackage{jheppub}
\usepackage{datetime}
\usepackage{float}
\usepackage{amsmath}
\usepackage{amssymb}
\usepackage{graphicx}
\usepackage{tikz}
\usepackage{multirow}
\usepackage{eufrak}
\usepackage{mathtools}
\usepackage{enumerate}
\usetikzlibrary{arrows,calc,positioning,shadows,shapes}

\graphicspath{ {./images/} }

\newcommand{\bea}{\begin{eqnarray}}
\newcommand{\eea}{\end{eqnarray}}
\newcommand{\be}{\begin{equation}}
\newcommand{\ee}{\end{equation}}
\newcommand{\nn}{\nonumber}
\newcommand{\dd}{\partial}
\newcommand{\ep}{\epsilon}

\begin{document}

\title{Instability of Black Holes in AdS$_3\times S^3$}

\author[a]{Finn Larsen}

\emailAdd{larsenf@umich.edu}

\author[b]{and Siyul Lee}

\emailAdd{siyul.lee@kuleuven.be}

\affiliation[a]{Leinweber Center for Theoretical Physics, University of Michigan, Ann Arbor, MI 48109, U.S.A.}
         
\affiliation[b]{Instituut voor Theoretische Fysica, KU Leuven,
Celestijnenlaan 200D, 3001 Leuven, Belgium}

\abstract{
Extremal and near extremal black holes are unstable, unless they are protected by supersymmetry. 
For black holes in AdS$_3 \times S^3$, we study the range of parameters that bound the unstable region, the fate of extremal black holes that are unstable, and the true ground state in super-selection sectors where no isolated black hole solution exists. Thermodynamics generally favors a central black hole surrounded by chiral primaries that carry a significant fraction of the conserved charges. We give a dynamical interpretation of decay, by computing the linear response functions in the dual CFT$_2$ and showing that the onset of instability agrees with thermodynamics. 
To understand the nature of the decay products, we construct several new families of classical string solutions. They are ``giants" that move on both the AdS$_3$ and the $S^3$, and may have interesting applications beyond those we develop.
}

\preprint{LCTP-25-09}

\maketitle


\section{Introduction}

The understanding of near extremal black holes has progressed significantly over the last few years. The linchpin of these advances is the existence of a near horizon geometry that is nearly AdS$_2$. Such a region is described by JT gravity and its dual Schwarzian theory. It is exciting that these theories can be analyzed reliably even in their strongly coupled quantum regime
\cite{Maldacena:2016hyu,Maldacena:2016upp,Stanford:2017thb,Iliesiu:2020qvm,Heydeman:2020hhw,Boruch:2022tno,Turiaci:2023wrh,Brown:2024ajk}.

However, in parallel with this development, it has become increasingly clear that all AdS$_2$ regions are unstable, unless they are protected by supersymmetry \cite{Durkee:2010ea,Bena:2011zw,Ezroura:2024xba}. Indeed, in many explicit cases, instability develops  strictly above the extremal limit, before an AdS$_2$ throat develops \cite{Kim:2023sig,Choi:2024xnv}.
 This situation is confusing, at the very least, and could undermine important results from the entire program. Therefore, improved understanding of black hole instabilities is essential for further progress. In this paper we study classical instability of extremal and near extremal black holes in AdS$_3\times S^3$, also known as BTZ black holes in supergravity with $(4, 4)$ supersymmetry. 
  
One might think that classical instability of a black hole is so dramatic that it is possible only in exceptional circumstances, but by now that impression appears incorrect. 
Some scenarios refer to black holes with ``hair" \cite{Bhattacharyya:2010yg,Aretakis:2011ha,Aretakis:2012ei,Murata:2012ct,Markeviciute:2016ivy,Horowitz:2022mly,Horowitz:2024kcx}.
We are inspired by the simple prescriptions developed recently in AdS$_{d+1}$ with $d\geq 3$ \cite{Kim:2023sig,Choi:2024xnv,Bajaj:2024utv,Choi:2025lck}.
In each case, a progenitor black hole decays to a central black hole remnant surrounded by some type of debris. The interaction between the two components of the composite is negligible, so the total of any conserved charge, including the energy, can be computed as the sum of two contributions. The decay is thermodynamically favored if the entropy of the final state, which is entirely carried by the black hole remnant, is larger than the initial state. Whenever this scenario is justified, simple thermodynamics identifies the true ground state in any super-selection sector defined by the conserved charges. There are several interrelated questions: 
\begin{enumerate}
\item
For generic conserved charges, there are {\it no} black holes that saturate the BPS bound for the mass. For charge configurations that do not permit a single-centered supersymmetric black hole, what is the gravitational representation of the supersymmetric ground state? 

\item
For given conserved charges, the black hole with the lowest possible mass is an extremal black hole with AdS$_2$ near horizon geometry. Generically, these black holes are unstable and we ask: what are the decay products? 

\item
For given conserved charges, what is the smallest energy for which the correct description of the system is a black hole with a single center? 
 \end{enumerate}
We pursue these questions in the context of AdS$_3\times S^3$. The first entry was previously studied in great detail, insofar as it can be identified 
with multi-center black hole solutions in asymptotically flat 4D spacetime and their wall-crossing phenomena \cite{Denef:2007vg,Bena:2011zw}, but the others are new. 

Before turning to the specifics of AdS$_3$ recall that, in supergravity with sufficient supersymmetry, unitarity imposes a BPS bound on the black hole hole mass that takes the linear form:\footnote{The mass $M$ and electric charges $Q_i$ are in units of the AdS-scale $\ell$, so they are pure numbers. Therefore, the angular momenta $J_a$ are scaled to zero in the asymptotically flat limit.}
\begin{equation}
M \geq\sum_a J_a + \sum_i Q_i = M_*~. 
\label{eqn:BPSboun} 
\end{equation}
The ranges of the indices $a$ and $i$ that enumerate the angular momenta $J_a$ and R-charges $Q_i$ depend on the number of spacetime dimensions and supersymmetries. The inequality \eqref{eqn:BPSboun} is saturated when $M=M_*$, and then the black hole preserves some supersymmetry. 
If a black hole has mass excess $M-M_*>0$ above the unitarity bound, this excess mass is unchanged upon emission of a particle that saturates the BPS bound by itself. 
Therefore, according to the first law of thermodynamics in the form: 
\begin{eqnarray}
T dS &=& dM - \sum_a \Omega_a dJ_a  - \sum_i \Phi_i dQ_i \cr
&=& d(M -  M_*)- \sum_a (\Omega_a -1)dJ_a  - \sum_i (\Phi_i -1)dQ_i ~,
\label{eqn:1stLaw} 
\end{eqnarray}
this process can increase the entropy if any $\Omega_a>1$ or $\Phi_i>1$. This criterion persists in the extremal limit where $T\to 0$. 
``Overrotating" black holes with $\Omega_a>1$ can emit particles that themselves satisfy $M=M_*$, by superradiance. 
Similarly, ``over-charged" black holes with $\Phi_i>1$ can emit charged particles.\footnote{The standard families of AdS black hole solutions do not include any examples where $\Omega_a>1$ or $\Phi_i>1$ at the same time. Because of this ``phenomenological" fact, the debris that ``dresses" the black hole has either angular momentum or electric charge, but never both.}

The AdS$_3$ example is special, because there are no propagating gravitons in bulk and the spacetime symmetry group factorizes into two chiral factors. We assume $(4, 4)$ supersymmetry, interpreted geometrically as descended from an $S^3$ factor. The theory has $SU(2)$-levels $k$ that are related to central charges through $c= 6k$. The energy-momentum quantum numbers are $(E, J_3)$ and the $R$-charges $(Q_L, Q_R)$. There is an independent BPS bound of the form \eqref{eqn:BPSboun} for each chirality but, without loss of generality, we can focus on the left,
\begin{equation}
\label{eqn:BPSbound}
E - E_{\rm SUSY} \geq J_3 + Q_L ~.
\end{equation} 
The supersymmetric Casimir energy is $E_{\rm SUSY} = - \frac{1}{2}k_L$ \cite{Assel:2015nca,ArabiArdehali:2018mil}. 
In a super-selection sector defined by charges $(J_3, Q_L, Q_R)$, the theory allows the continuous range of energies $E$ given by \eqref{eqn:BPSbound}.\footnote{Because of quantum effects, the spectrum of allowed energies has a small gap between the BPS bound and the continuum \cite{Iliesiu:2020qvm,Heydeman:2020hhw,Boruch:2022tno}. This detail does not play any role in our study.}

For energies $E$ that satisfy 
\begin{equation}
\label{eqn:extbound}
E \geq E_{\rm ext} = J_3 + \frac{1}{2k_L}Q^2_L ~, 
\end{equation} 
there is a single center black hole with the given charges. The extremality bound \eqref{eqn:extbound} is at least as demanding as the BPS bound \eqref{eqn:BPSbound}. They can be
simultaneously saturated only if $Q_L = k_L$. In this case the extremal black hole is supersymmetric and has entropy: 
\begin{equation}
S_{\rm BPS} = 2\pi \sqrt{k_R \left( J_3+\frac{1}{4}k_L \right) - \frac{1}{4}Q^2_R}~.
\end{equation}
In a super-selection sector defined by $(J_3, Q_L, Q_R)$ with $Q_L\neq k_L$ we ask: 
\begin{enumerate}
\item
What $E$ is sufficiently large that a single center black hole dominates thermodynamically?
\item
Extremal black holes saturate \eqref{eqn:extbound} and are unstable, but what do they decay to? 
\item
For $E$ that saturate the BPS bound \eqref{eqn:BPSbound}, what is the gravitational representation of 
the quantum state? 
\end{enumerate}
It is sufficient to study charges in the range $0\leq Q_L \leq 2k_L$. Super-selection sectors outside this range can be reached by 
spectral flow \cite{Schwimmer:1986mf,Maldacena:1998bw,Kraus:2006nb}.

The excitations that saturate the supersymmetric bounds are the {\it chiral primaries} \cite{deBoer:1998kjm,deBoer:1998us}, a lead central character in all supersymmetric CFT$_2$'s. Therefore, our baseline hypothesis is that, when considering the questions enumerated above, the dominant competitor to a single center black hole is a central black hole dressed by a chiral primary. Physically, ``the" chiral primary could be realized by multiple single particle gravitons, but this would be inconsequential. We further assume that, in the classical approximation, the quantum numbers of the composite system is the sum of its components, and its entropy is entirely carried by the central black hole. Accordingly, the central BTZ black hole is not necessarily supersymmetric, but the ``gas" particles are. 

According to our hypothesis, in a super-selection sector where no supersymmetric black hole exists, the BPS ground state corresponds to supersymmetric black hole center dressed by a chiral primary.\footnote{Such configurations were called ``moulting" black holes \cite{Bena:2011zw} and the mechanism was called the ``dressed concentration conjecture" in  \cite{Choi:2025lck}.
``Enigmatic" examples are more general in that several centers may have classical entropy \cite{Denef:2007vg}. } Such supersymmetric ground states were studied using multi-center BPS black hole in four asymptotically flat dimensions \cite{Bates:2003vx}. As such, they are closely related to the wall-crossing phenomenon, which can be identified also in the dual symmetric orbifold CFT$_2$. These results, and the techniques that established them, rely heavily on supersymmetry. Our main interest is to study the analogous questions  for non-supersymmetric black holes. 

Even though we abandon supersymmetry, we have a high degree of confidence in our result that many black holes in AdS$_3\times S^3$ are susceptible to disintegration. 
To be sure, BTZ black holes always have potential $|\Omega| \leq 1$ for rotation within AdS$_3$, so they are stable against superradiant emission within AdS$_3$. However, the potentials for R-charge, interpreted as rotation within $S^3$, 
can have $\Phi_{L,R}>1$.\footnote{AdS$_3\times S^3$ is the near horizon limit of a 6D black string. Excitations of this string are related to microstates of the 5D BMPV black hole. In this context, the conventional notation for quantum numbers is $(P, J_L, J_R)$ where $P$ is momentum along the KK-direction of the 6D string, and the angular momenta $J_L, J_R$ are charges corresponding to rotation along $S^3$. In this article we take an AdS$_3$ perspective inspired by AdS$_{d+1}$ with $d>2$ and then the notation $(J_3, Q_L, Q_R)$ is appropriate.} In this situation the first law  \eqref{eqn:1stLaw} shows that emission lowering $Q_{L,R}$ is thermodynamically advantageous. That gives robust evidence that the low energy regime is controlled by a charged condensate.\footnote{The AdS$_3\times S^3$ instability is analogous to ``dual dressed" black holes in higher dimensions \cite{Choi:2024xnv}
but distinct from the ``grey galaxy" and ``revolving black hole" varieties \cite{Kim:2023sig}.}

The dynamics underlying the instability of black holes in AdS$_3\times S^3$ is reminiscent of superradiance, but adapted to the AdS-boundary conditions. To show how, we compute the retarded Green's functions for general conformal fields and interpret them in terms of linear response theory. Unstable black holes experience negative friction that pulls towards an equilibrium with a nontrivial condensate. The instability identified this way has onset precisely where indicated by thermodynamics. 

Thermodynamically, the most favored decay is via emission of chiral primaries. More precisely, the very most favored process would be to emit (chiral, chiral) primaries. Such a process would leave a central black hole with angular momenta on the $S^3$ completely fixed at $Q_L=k_L$ and $Q_R=k_R$. For example, the supersymmetric BMPV black holes can increase their entropy, which includes a $-Q_L^2$ under a square-root, by lowering the angular momentum $Q_L$. This conclusion is suggested by thermodynamics, but most likely it cannot be realized in a UV complete theory. The standard symmetric orbifold theories do have (chiral,chiral) primaries with $E_L=Q_L$, $E_R=Q_R$ but their spin $Q_L-Q_R$ cannot be large, it does not increase in the classical limit. For example, the Poincar\'{e} polynomial of $K3^k/S_k$ does not generate states with spin $\geq 2$, even when $k$ is large \cite{deBoer:1998kjm,deBoer:1998us}.

To make progress we need details of the spectrum in a regime where it is not well understood. A central question is the scope of the stringy exclusion principle \cite{Maldacena:1998bw}. In the symmetric orbifold CFT$_2$ \cite{Strominger:1996sh}, a chiral 
primary with large charge fills out a finite fraction of the target space. This leaves a significantly smaller effective central charge for the CFT$_2$ sector responsible for the black hole entropy \cite{Bena:2011zw,Martinec:2023plo}. This is entropically disfavored and should be avoided, if possible. 

It is instructive to compare with AdS$_5\times S^5$ which is dual to ${\cal N}=4$ SYM. In that context $1/2$-BPS states are in one-to-one correspondence with Young tableaux of the dual $SU(N)$ gauge theory. The height of columns in the Young tables are limited by $N$ and correspond to $N$ identical operators that are fully antisymmetrized in the gauge indices. Classically, such operators are interpreted as giant gravitons, equivalent to D3-branes wrapped on cycles that are $S^3\subset S^5$. However, there is no limit on the charge of dual giants that wrap cycles within $S^3\subset$AdS$_5$, because this corresponds to the width of the Young tableaux. Dual giants are thermodynamically favored because the stringy exclusion principle does not apply, in the sense that they can carry charges $\sim N^2$ without reducing the effective radius of AdS$_5$ \cite{Choi:2024xnv,Choi:2025lck}. 

In our thermodynamic study we do not commit to whether the stringy exclusion principle applies in AdS$_3$, we consider both options because details depend on assumptions about the UV theory. However, motivated by this question, and by the phases identified in higher dimensions, we are led to seek excitations that are localized on $S^3$. We consider a ``giant" string that responds to both the fluxes that support AdS$_3$ and the dual $S^3$. We find classical solutions supported by rotation in both AdS$_3$ and $S^3$. Such mixtures of giants and dual giants are novel, they do not have known analogues in higher dimensions. 

Generally, the class of giant string solutions we construct  are not supersymmetric but, in certain limits, they can preserve $\frac12$
or $\frac14$ of the supersymmetries. The latter are interesting because each of these strings can carry macroscopic charges, comparable to that of the black hole. Since they carry large charge, without much flux, they effectively evade the stringy exclusion principle. This makes their thermodynamics interesting. 

This article is organized as follows. 
\begin{itemize}
\item
In Section \ref{sec:thermo}, we discuss the basic thermodynamics of phase separation. We assume that, in the classical approximation, the entropy is entirely carried by the remaining black hole, the entropy of matter is negligible, and we also neglect interactions between the central black hole and its decay products. 
We find that BTZ black holes can increase entropy by emitting matter while conserving all applicable charges, including the energy. 

\item
In Section \ref{sec:response} we discuss the dynamical instability, by examining a probe field in the black hole background. The AdS mechanism is the analogue of superradiance
for asymptotically flat black holes. We show that, exactly when the thermodynamic instability applies, matter with the charges of a chiral primary can be repelled from the black hole.
The instability can be interpreted as negative friction in the dual theory and its onset is associated with a superfluid condensate.

\item
In Section \ref{sec:giantstrings}, we construct several new families of giant strings with many parameters that may well be interesting for other purposes. These giant strings rotate along great circles in both AdS$_3$ and $S^3$, and they are localized in both the radial direction of AdS$_3$ and the polar angle of $S^3$. They are spectators, no backreaction is taken into account. 

\item
In Section \ref{sec:giantthermo} we develop the thermodynamics of the giant strings.
We re-examine the thermodynamic principles from Section \ref{sec:thermo}
under the assumption that a small number of giant strings are allowed along with a black hole in the core.
This composite system exhibits several thermodynamic phases.

\item
We conclude in Section  \ref{sec:discussion}, with a discussion of open problems. In this setting we comment on related recent research. 

\item
In Appendix \ref{sec:reviewGS} we review the construction of supersymmetric giants and dual giants in AdS$_3\times S^3$. 
\end{itemize}

\section{Instability from Thermodynamics}\label{sec:thermo}

In this section we study instability of BTZ black holes against emission of particles.
We first review the thermodynamic properties of the black hole, and then identify 
a simple scenario where the black hole gains entropy by emission of particles that are chiral primaries.

\subsection{The AdS$_3$ Black Hole and its Stability}

The BTZ black holes we study derive from rotating black hole solutions \cite{Cvetic:1996xz}
to $\mathcal{N}=4$ or $\mathcal{N}=8$ supergravity in 5D,
by interpreting it as black strings in 6D and taking the decoupling limit \cite{Cvetic:1998xh}.
Locally, the resulting black hole geometry is a direct product of a BTZ black hole \cite{Banados:1992wn,Banados:1992gq}
to AdS$_3$ gravity and a three-sphere $S^3$ with equal but opposite constant curvatures.
The global structure encodes rotation of the $S^3$ with respect to the time in AdS$_3$.

When interpreted as a 3D spacetime, the BTZ black hole  becomes a solution to AdS$_3$ supergravity, with two $SU(2)$ gauge fields 
taking the $S^3$ into account. The black hole is parametrized by four conserved quantities: the energy (or mass) $E$ and the angular momentum $J_3$
that come from the $SO(2,2)$ isometry  of AdS$_3$, and two charges $Q_L$ and $Q_R$ due to the $SO(4) \sim SU(2)_L \times SU(2)_R$ isometry of $S^3$.
We will also use the chiral linear combinations $E_R = E+J_3$ and $E_L = E-J_3$ and we refer to all four conserved quantities $(E_R,E_L,Q_R,Q_L)$ 
simply as ``the" charges. Our conventions for charges and their conjugate potentials 
are summarized succintly by the first law of black hole thermodynamics:
\bea\label{1stlaw}
\frac{1}{\beta} dS_{\rm BH} = dE - \Omega dJ_3  - \Phi_R dQ_R - \Phi_L dQ_L~.
\eea
The relation between charges and thermodynamic potentials are:\footnote{Henceforth
we take the two $SU(2)$ levels equal:$k_L = k_R = k$.}
\bea\label{chargeaspot}
E_R ~\equiv~ E+J_3 &=& \frac{2 k}{\beta^2(1-\Omega)^2} \left( \pi^2 + \beta^2 \Phi^2_R\right)~, \nn\\
E_L ~\equiv~ E-J_3 &=& \frac{2k}{\beta^2(1+\Omega)^2} \left( \pi^2 + \beta^2 \Phi^2_L\right)~, \nn\\
Q_R &=& \frac{2k}{1-\Omega}  \Phi_R~, \nn\\
Q_L &=& \frac{2k}{1+\Omega}  \Phi_L~.
\eea
The physical ranges for the potentials are $-1 \leq \Omega \leq 1$ and $\beta^{-1} \geq 0$, in order that the conserved charges are finite.
Also, we restrict our attention to $0 \leq Q_{R/L} \leq 2k$, because the black holes
outside of this range are related to those inside by spectral flow \cite{Maldacena:1998bw,Kraus:2006nb}. 
The black hole entropy is 
\bea\label{BHentropy}
S_{\rm BH} &=& 2 \pi \sqrt{\frac{k E_{R}}{2} - \frac{Q_{R}^2}{4}}+
2 \pi \sqrt{\frac{k E_{L}}{2} - \frac{Q_{L}^2}{4}}~.
\eea
The relations \eqref{chargeaspot} and the formula \eqref{BHentropy} for the entropy satisfy the first law \eqref{1stlaw}.

Physical conditions impose two sets of inequalities on the black hole charges.
First, combinations of \eqref{chargeaspot} show
\bea\label{extbound}
E_R -  \frac{Q_R^2}{2k} = \frac{2\pi^2 k}{\beta^2 (1-\Omega)^2} &\geq& 0~, \nn\\
E_L -  \frac{Q_L^2}{2k} = \frac{2\pi^2 k}{\beta^2 (1+\Omega)^2} &\geq& 0~.
\eea
These inequalities are the extremality bounds, because the black hole temperature vanishes if and only if one of them is saturated.
Physically, they are required in order for the black hole in Lorentzian signature to have regular horizon. 

The second set of inequalities are the BPS bounds:
\bea\label{BPSbound}
E_R &\geq& Q_R - \frac{k}{2} ~, \nn\\
E_L &\geq& Q_L - \frac{k}{2} ~.
\eea
In supergravity, these inequalities are required by unitarity. 
Saturation of either implies preservation of $\frac14$ of the supersymmetries. This is enhanced
to $\frac12$ if both are saturated. A BPS bound (\ref{BPSbound}) is automatically \emph{satisfied} if the corresponding extremality condition (\ref{extbound}) is satisfied.
On the other hand, a BPS bound (\ref{BPSbound}) can be \emph{saturated}
only when the corresponding one in (\ref{extbound}) is also saturated,
and that happens precisely when
\be\label{AdS3cc}
Q_{R/L} = k~.
\ee
A regular and supersymmetric black hole that saturates (\ref{BPSbound})
must have \eqref{AdS3cc} in addition to $E_{R/L} = \frac{k}{2}$, and thus be extremal.
In Lorentzian signature, supersymmetric black holes are always extremal, but the converse is not automatic. 
It is a universal feature of supersymmetric AdS black holes in all dimensions
that is not fully understood microscopically \cite{Larsen:2021wnu,Larsen:2024fmp}.

Our interest is the stability of the black hole towards emission of particles. A candidate for emission has 
quantum numbers $(\ep,j_3,q_R,q_L)$ that are much smaller than the macroscopic charges of the black hole.
The black hole charges decrease accordingly, and the first law (\ref{1stlaw}) gives the change in its entropy: 
\be\label{DeltaSgeneral}
\frac{1}{\beta} \Delta S_{\rm BH} = -\ep + \Omega j_3  + \Phi_R q_R + \Phi_L q_L~.
\ee
If this quantity is positive, the black hole gains entropy in the process. 
Therefore, provided that emission is dynamically allowed, the black hole is unstable against the decay. In the following, 
we study this criterion using various assumptions about the particles that are available for the decay. 

\subsection{Emission of Chiral Primaries}\label{sec:scenariogeneric}

To determine the most entropic configuration in a super-selection sector defined by the total conserved charges, 
we must make assumptions about the matter content of the theory. Matter particles must, at a minimum, satisfy 
the BPS bounds:
\bea\label{BPSforparticle}
e_R \equiv \ep + j_3 \geq |q_R|~, \qquad
e_L \equiv \ep - j_3 \geq |q_L|~.
\eea
Note that we use lowercase symbols for charges of the gas particles. In this subsection we do not take backreaction into account so we simply add matter charges, including the energy, to those of the black hole. We also assume that the entropy of the particles is negligible compared to that of the black hole,
so the entropy of the system equals that of its black hole component.

Let $E_{R/L}$ and $Q_{R/L}$ be the total charges of the system.
They must satisfy the BPS bounds \eqref{BPSbound}, {\it wiz.} $E_{R/L} \geq Q_{R/L} - \frac{k}{2}$.
Each charge is separated into contributions from the black hole and from the particles so that
\be\label{distributecharges}
E_R = E_{R,BH} + e_R~,
\ee
and so on.
By assumption, the entropy of the system is dominated by that of the black hole:
\bea\label{BHentropyV2}
S_{BH} &=& 2 \pi \sqrt{\frac{k E_{R,BH}}{2} - \frac{Q_{R,BH}^2}{4}}+
2 \pi \sqrt{\frac{k E_{L,BH}}{2} - \frac{Q_{L,BH}^2}{4}}~.
\eea
The goal is to maximize (\ref{BHentropyV2}), given total charges $E_{R/L}$ and $Q_{R/L}$ that are distributed between the black hole and the particles as in \eqref{distributecharges}.
Both for the black hole and for the particles, the right- and left-charges
i.e. $(E_R,Q_R)$ and $(E_L,Q_L)$ are not correlated.
Therefore we can maximize each of the two terms in (\ref{BHentropyV2}) independently.

\begin{figure}
\begin{center}
\includegraphics[width=0.65\textwidth]{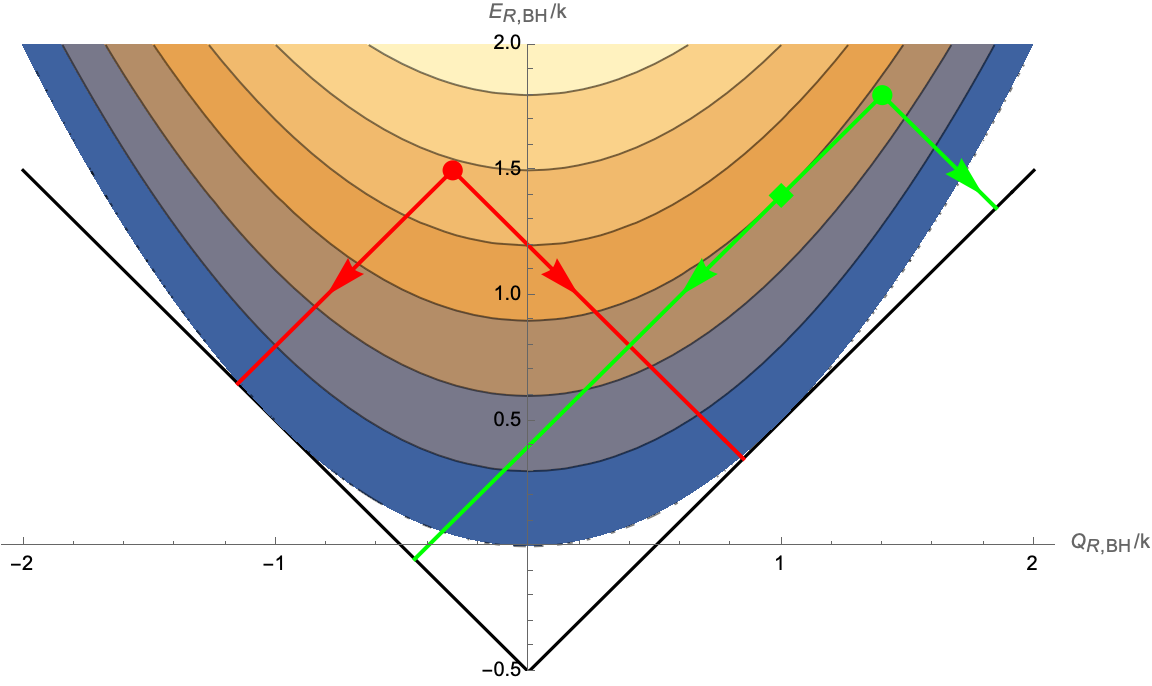}
\caption{\label{Fig:ERJRplane} The black hole charge $(Q_{R,BH},E_{R,BH})$-plane
in units of $k$.
Colors indicate black hole entropy, the yellower the higher.
Given a black hole at any point of the diagram, candidate decays that conserve charges
are towards the region bounded by two lines going down at $\pm 45^\circ$ angles. 
For a black hole at the green circle, the decay that maximizes the entropy is at the green diamond to its SW.
For a black hole at the red circle all possible decays reduce the entropy, so this black hole is stable.}
\end{center}
\end{figure}

We first study the side involving total charges $(E_R,Q_R)$ such that $E_{R} > Q_{R} - \frac{k}{2}$. 
We seek the $e_R$ and $q_R$ that satisfy $e_R \geq |q_R|$ and maximize the entropy of the central black hole: 
\be\label{BHentropyR}
S_{BH,R} = 2 \pi \sqrt{\frac{k (E_R-e_R)}{2} - \frac{(Q_R-q_R)^2}{4}}~.
\ee
This can be answered algebraically, but perhaps more easily using a graphic.
In Figure \ref{Fig:ERJRplane}, the shadings indicate the black hole entropy $S_{BH,R}$
as a function of $(Q_{R,BH},E_{R,BH})$, with yellow corresponding to higher entropy.
Note the absence of black holes below the parabolic extremality bound \eqref{extbound}.
The red and green circles represent examples of the total charges $(Q_R,E_R)$.
Because of the bounds $e_R \geq |q_R| \geq 0$ on the matter, 
the range of possible black hole charges $(Q_{R,BH},E_{R,BH})$ in each example
are represented by the region between the two diagonal lines that emanate from the circle downwards, but within the shaded region for the black hole to exist.

The green circle corresponds to the total charges $(Q_R,E_R) = (1.4k, 1.8k)$. Within the allowed region for the black hole charges, the 
R-entropy \eqref{BHentropyR} is maximized  at the point $(Q_{R,BH},E_{R,BH}) = (k,1.4k)$, represented by the green diamond.
Thus, given the specified total charges, the most entropic configuration consists of a black hole with charges $(Q_{R,BH},E_{R,BH}) = (k,1.4k)$,
and particles that carry the remaining charges $(q_R,e_R) = (0.4k, 0.4k)$.
Consequently, a black hole with the total charges $(Q_{R,BH},E_{R,BH}) = (1.4k,1.8k)$,
is unstable against particle emission. The particles emitted in the optimal process satisfy $q_R = e_R$, so they are chiral primaries in the CFT$_2$ language.

The red circle corresponds to total charges $(Q_R,E_R) = (-0.3k, 1.5k)$.
In this case the allowed region for black hole charges is between the two red diagonal lines. However, the 
R-entropy \eqref{BHentropyR} is not lowered anywhere in the region, it is maximized at the red circle.
Thus, a black hole with such total charges is thermodynamically stable against emitting any particles that obey \eqref{BPSforparticle}.

The boundary that distinguishes the two qualitatively different phases,
defined by whether the entire charge should be carried by a black hole,
are the vertical lines $|Q_R| = k$.
If $\pm Q_R > k$, the most entropic configuration is such that an appropriate (anti-)chiral primary with $e_R = \pm q_R$ is emitted,
and the remnant black hole carries exactly $Q_{R,BH} = \pm k$.
If on the other hand $|Q_R| < k$, then it is entropically favored for the black hole
to carry all charges of the system in their entirety, leaving no room for any other particles.

All arguments are symmetric under $R \leftrightarrow L$, so the analogous discussion applies independently to the distribution of the total charges $(E_L, Q_L)$
between the black hole and particles. For example, if $0 < Q_R < k$ and $k < Q_L$, then the entropy is maximized by the black hole that carries $E_R$ and $Q_R$ in their entirety,
but only $Q_{L,BH} = k$. Therefore, in this case, the most entropic configuration would be a black hole with
\be
E_{R,BH} = E_R~, \qquad Q_{R,BH} = Q_R~, \qquad
E_{L,BH} = E_L - (Q_L-k)~, \qquad Q_{L,BH} = k~,
\ee
coexisting with chiral primary particles that carry $e_L = q_L = Q_L-k$.

An important general lesson is that the most entropic configuration never involves a particle that is \emph{not} (anti-)chiral primary,
i.e. a particle with either $e_R \neq |q_R|$ or $e_L \neq |q_L|$. In other words, the matter component in the favored configurations must have 
$e_R = q_R$ \emph{and} $e_L = q_L$, with the possibility that one or both can vanish. Such matter is $\frac12$-BPS, and it is preferred because a particle 
with excess energy $e_{R/L} > |q_{R/L}|$ ``wastes" entropy. Black holes are the most efficient carriers of entropy, so it is favorable that they 
carry any energy that the matter is not required to have by its charge content.

The chiral primaries are the preferred debris also for systems whose total charges violate the extremality bound \eqref{extbound} but still obey the BPS bound \eqref{BPSbound}.
Crucially, such systems include supersymmetric ones that do not obey the supersymmetric black hole charge constraint \eqref{AdS3cc}, $Q_{R,BH} = k$.
For example, when the system has total charges $E_R + k/2 = Q_R > k$,
it cannot be realized by a regular, single-centered black hole solution.
However, the arguments of this subsection suggest a composite system that consists of
the BPS black hole with $E_{R,BH} = k/2$ and $Q_{R,BH} = k$,
and the required amount of chiral primaries $e_R = q_R = Q_R - k$.
Provided that the assumption of non-interaction is upheld,
the composite system serves as a candidate for supersymmetric ground states
with generic charges in AdS$_3$ supergravity \cite{Bena:2011zw}.

The arguments of this subsection provides a simple picture of AdS$_3$ black hole instability
and shows the thermodynamic preference of BPS particles, or (anti-)chiral primaries, being emitted by the black hole. However, there is evidence from the dual CFT$_2$ that chiral primaries with large but arbitrary charges
$q_R$ and $q_L$ are not available \cite{deBoer:1998kjm,deBoer:1998us}
(see also section 5.3.3 of the review \cite{Aharony:1999ti}).
Instead, the two charges must approach each other $q_R \approx q_L$ as they become large. Thus 
scenarios of thermodynamic (in)stability, such as the one discussed in this section, depend on the detailed spectrum of particles available to emission.
For this reason, we revisit the black hole stability problem in section \ref{sec:giantthermo},
after having constructed some general classes of giant string solutions in the
AdS$_3 \times S^3$ background in section \ref{sec:giantstrings}.

\subsection{Emission of Non-Rotating Particles}

In this subsection, we study instability of AdS$_3$ black holes against emission of particles that do not rotate ($j_3 = 0$), but
carry one of the angular momenta in $S^3$,
say $q_L > 0$ and $q_R = 0$.
We take them to be $\frac14$-BPS, so $\epsilon = q_L$.
The Type II giant string solutions discussed in section \ref{sec:BBsoln} realize such objects with any $\epsilon = q_L$ but in this subsection we do not require any details about what they are. 

The condition for (in-)stability of an AdS$_3$ black hole against emission of non-rotating particles can be easily inferred from the first law \eqref{DeltaSgeneral}.
The black hole is stable if its electric potential is such that
\be\label{Phi1bound}
\Phi_L \leq 1~,
\ee
which is the universal criterion for black hole stability.
This criterion can be written in terms of the black hole charges
by inverting the relations \eqref{chargeaspot} between charges and potentials. It is equivalent to
\be\label{eqn:Eboundgeneral}
\frac{Q_L}{k} - 1 \leq \frac{\sqrt{E-J_3-\frac{Q_L^2}{2k}}}{\sqrt{E+J_3-\frac{Q_R^2}{2k}}}~.
\ee
The expressions inside square roots are non-negative, by
the black hole extremality bounds \eqref{extbound}.

\begin{figure}[h]
\centering
\includegraphics[width=0.8\linewidth]{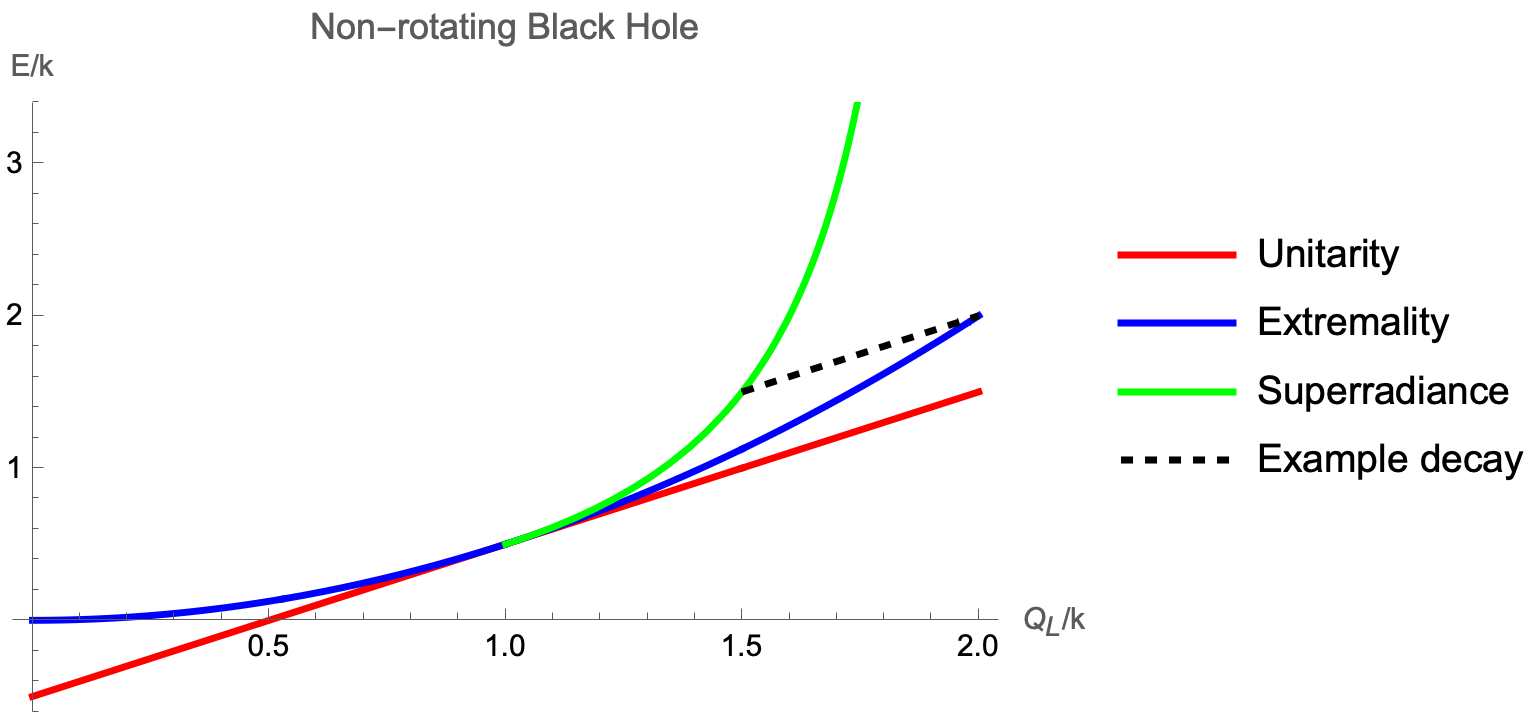}
\caption{Non-rotating BTZ black holes are thermodynamically stable for energies
above the superradiance bound and for $Q_L \leq k$.
For lower energies -- but not below the extremality bound -- black hole solutions exist, but they are unstable. 
At even lower energies, between the extremality and unitarity bound, no candidate black hole exists. The dashed line gives an example of an extremal black hole that decays and ends at the superradiance bound.}
\label{fig:bounds}
\end{figure}

To simplify the presentation, we focus on non-rotating black holes ($J_3 = 0$)
that carry only one of electric charges ($Q_L > 0$, $Q_R = 0$).
Then \eqref{eqn:Eboundgeneral} is automatically satisfied if $Q_L \leq k$ and, if $Q_L > k$, it reduces to
\be\label{eqn:Ebound}
\frac{E}{k}\geq \frac{1}{2 ( \frac{2k}{Q_L}-1)}~.
\ee
Since
$$
\frac{1}{2 ( \frac{2k}{Q_L}-1)} \geq \frac{1}{2k}Q^2_L \quad\Leftrightarrow\quad
\left( \frac{Q_L}{k}-1\right)^2 \geq 0~,
$$
the new bound is more strict than extremality, as expected.
Any black hole with $Q_L > k$ that satisfies the superradiance bound 
also satisfies the extremality bound,
but there are examples of regular black holes that violate the superradiance bound.
The bounds are illustrated in Figure \ref{fig:bounds}.

For definiteness, we focus on the fate of an extremal black hole. By definition, it saturates \eqref{extbound}, so $E_{\rm tot} = \frac{Q^2_{\rm tot}}{2k}$. If $1<\frac{Q_{\rm tot}}{k}\leq 2$, it is unstable towards a black hole remnant and a giant string. The central black hole has energy  $E_{\rm BH}$ and charge $Q_{\rm BH}$ that saturate the superradiance bound \eqref{eqn:Ebound}: 
\begin{equation}
\frac{E_{\rm BH}}{k} = \frac{Q_{\rm BH}}{2(2k-Q_{\rm BH})}~.
\label{eqn:supradbound}
\end{equation}
The two components of the final state do not interact, so the total charges are conserved: 
$E_{\rm tot} = E_{\rm BH} +e$ and $Q_{\rm tot} = Q_{\rm BH} + q$. This gives
\be
Q_{\rm tot} - Q_{\rm BH} = q = e= E_{\rm tot} - E_{\rm BH} = \frac{Q^2_{\rm tot}}{2k} - \frac{kQ_{\rm BH}}{2(2k-Q_{\rm BH})}~.
\ee
It can be solved for $Q_{\rm BH}$ to yield
\begin{align}
\frac{Q_{\rm BH}}{k}  & =\frac{1}{4} \left(  3 + \frac{2Q_{\rm tot}}{k} -\frac{Q^2_{\rm tot}}{k^2}\right) + \frac{1}{4}
\Big(\frac{Q_{\rm tot}}{k} -1\Big) \sqrt{ \frac{Q^2_{\rm tot}}{k^2} - \frac{2Q_{\rm tot}}{k} + 9 }~.
\end{align}
The details of this expression are not particularly illuminating.
Its significance is in showing that the charge of the central black hole in the final state is in the range $1\leq \frac{Q_{\rm BH}}{k}\leq \frac{Q_{\rm tot}}{k} $. It has less charge than the progenitor, but more than a BPS black hole. 

The initial extremal black hole has entropy
\be
S_{\rm tot} = 2\pi\sqrt{ \frac{1}{2}kE_{\rm tot}} = \pi Q_{\rm tot}~.
\ee
In the final state, the entire entropy is carried by the black hole remnant:
\be
S_{\rm BH} = 2\pi\sqrt{ \frac{1}{2}kE_{\rm BH}}  + 2\pi\sqrt{ \frac{1}{2}kE_{\rm BH}-\frac{1}{4}Q^2_{\rm BH}}
=  \pi  \sqrt{ \frac{Q^3_{\rm BH}}{2k -Q_{\rm BH}}}~.
\ee
This value is larger than the initial value. 

The special case of $E_{\rm tot}=2k$ and $Q_{\rm tot}=2k$ involves particularly simple numbers.
Even though $E_{\rm tot}=Q_{\rm tot}$, this black hole is not supersymmetric.
Because of the Casimir energy $E_{\rm SUSY} = - \frac{1}{2}k$,
the condition for a supersymmetric black hole was $E_{\rm BPS}= Q_{\rm tot} - \frac{k}{2}$ \eqref{BPSbound}.
As it is, the initial black hole is extremal, $E_{\rm tot} = \frac{1}{2k}Q^2_{\rm tot}$. The decay of this extremal black hole ends with a central black hole that has $E_{\rm BH} = \frac{3}{2}k$ and $Q_{\rm BH} = \frac{3}{2}k$.
These values saturate the bound \eqref{eqn:Ebound}, so this black hole is maximally spinning and on the verge of superradiant instability.
The other decay product is a giant string with $E_{\rm giant} = Q_{\rm giant} = \frac{1}{2} k$. 
In this process, the entropy increases from $2\pi k$ to $\frac{3\sqrt{3}}{2} \pi k\sim 2.599\pi k$. The decay with these parameters is plotted as a dashed line on Figure \ref{fig:bounds}.

\section{Dynamics of Instability: Linear Response Theory for BTZ}
\label{sec:response}

In this section we discuss the dynamics of the instability of black holes in AdS$_3\times S^3$. This complements 
the thermodynamic analysis presented in the previous section. 

\subsection{The Superradiant Instability}
\label{subsec:superradiant}

The dynamical mechanism is ultimately a version of the superradiant instability
that is familiar from Kerr black holes in asymptotically flat 4D spacetime
(see \cite{Brito:2015oca} and references therein).
In that context, and similar ones, the emission rate of Hawking radiation takes the form: 
$$
\Gamma(\epsilon,j_3) \frac{d^3k}{(2\pi)^3}= \frac{\sigma_{\rm abs}(\epsilon,m_z)}{e^{\beta(\epsilon - j_3\Omega)} -1} \frac{d^3k}{(2\pi)^3}~.
$$
The energy $\epsilon$ and the third component of the angular momentum $j_3$ are parameters of the emitted particle. 
The inverse temperature $\beta$ and the rotational velocity $\Omega$ refer to the Kerr black hole. 
The absorption cross-section $\sigma_{\rm abs}(\epsilon,j_3)$ encodes scattering in the black hole geometry. The important feature is that the emission 
rate diverges at sufficiently low energy $\epsilon\to j_3\Omega$ and then becomes negative as $\epsilon< j_3\Omega$, at least formally. This is the superradiant instability.\footnote{
We present the effect in terms of spontaneous emission of Hawking radiation for conceptual clarity, but we might as well interpret it as stimulated emission: radiation that is incident on the black hole reflects back with enhanced amplitude. This aspect of superradiance is entirely classical.} 

Black holes in asymptotically flat space are somewhat artificial, because they have negative specific heat. Therefore, the ground state at any temperature, or any fixed energy, is a widely dispersed dilute gas. This conclusion can be circumvented by regulating infrared behavior. i.e. by studying the black hole in a box. In the modified setup emitted radiation can reflect from the boundary at infinity and return to the black hole. This mechanism gives a runaway instability and identifies the physical ground state as a black hole in equilibrium with a dilute gas. 

We have reviewed the 4D Kerr black hole example because it introduces several important features. There is a dependence on the spectrum of the theory in that superradiance applies to particles with $\epsilon - j_3\Omega<0$. For the 4D Kerr, gravitational perturbations with long wavelength always satisfy this condition, but gravity waves have suppressed absorption cross section $\sigma_{\rm abs}$ at low energy, so the balance requires a detailed study. The Kerr example also introduces the importance of boundary conditions, which can determine the qualitative behavior. 

In AdS space-times, the boundary conditions are given by the geometry, and they implement a physical box. However, the spectrum depends on additional assumptions. As discussed in Section \ref{sec:thermo}, AdS$_3$ thermodynamics suggests that there is an instability when the black hole potentials $(\Omega, \Phi_L, \Phi_R)$ and quantum numbers of matter in the theory $(\epsilon, j_3, q_L, q_R)$ are such that
\be
\epsilon - \Omega j_3 - \Phi_R q_R - \Phi_L q_L 
= (1 + \Omega) \frac{ \epsilon-j_3}{2} + (1 - \Omega) \frac{ \epsilon+j_3}{2} - \Phi_R q_R - \Phi_L q_L < 0~.
\ee
For example, right- or left-chiral matter can satisfy the instability condition if
\be
(1 + \Omega) \frac{ \epsilon-j_3}{2} - \Phi_R q_R < 0~, \qquad
(1 - \Omega) \frac{ \epsilon+j_3}{2} - \Phi_L q_L < 0~,
\label{eqn:normalocn}
\ee
respectively.
The relation to the condition $\epsilon-\Omega j_3 < 0$ that controls superradiance is suggestive, but the physics of asymptotically flat space is very different from AdS. 
In the following, we discuss a variant of the superradiat instability that applies in AdS. 

\subsection{Effective Potential for Motion in the BTZ Background}
\label{subsec:effpot}

The BTZ geometry is
\begin{equation}
\label{eqn:BTZbh}
ds^2 = - \frac{(r^2 - r^2_+)(r^2 - r^2_-) }{r^2\ell^2} dt^2 +  \frac{r^2\ell^2}  {(r^2 - r^2_+)(r^2 - r^2_-) } dr^2 + r^2  (d\phi + \frac{r_+ r_-}{\ell r^2} dt)^2 ~. 
\end{equation}
The black hole parameters enter through the coordinate locations $r_\pm$ of the inner and outer horizons: 
\begin{eqnarray}
\label{eqn:Mrpm}
E \pm J_3 &=&  k \frac{(r_+ \pm  r_-)^2}{2\ell^3}~.
\label{eqn:Jrpm}
\end{eqnarray}
Here $k = \frac{1}{6} c = \frac{\ell}{4G_3}$ is the (common) level of the $SU(2)$ currents of the $(4,4)$ supersymmetry. 

The Klein-Gordon equation for a massive particle in the BTZ background \eqref{eqn:BTZbh} reduces to the radial equation
\begin{equation}
\frac{1}{r} \partial_r \left( \frac{(r^2-r^2_+)(r^2-r^2_-)}{r}\partial_r \right)\psi(r) = V_{\rm eff} \psi(r)~,
\label{eqn:KGeqn}
\end{equation}
for a wave function of the form $\Psi = \psi(r) e^{i(\epsilon t - j_3 \phi)}$. The effective potential: 
\begin{equation}
V_{\rm eff}(r) = \frac{1}{r^2_+-r^2_-} \left(  - \frac{ ( \epsilon r_+  - j_3 r_-)^2} {r^2-r^2_+} + \frac{(\epsilon r_- - j_3  r_+ )^2 } {r^2-r^2_-}  \right) + m^2~,
\label{eqn:Veff}
\end{equation}
expresses the dynamics succinctly. It depends on the black hole parameters through $r_+, r_-$ and the probe has parameters $\epsilon, j_3, m$.   

The effective potential $V_{\rm eff}(r)$ can be interpreted in terms of particle motion in the region outside the black hole $r>r_+$. 
It has several notable features: 
\begin{itemize}
\item
The mass term proportional to $m^2$ {\it screens} asymptotic infinity, preventing particles with mass $m^2>0$ from reaching the conformal boundary at 
asymptotic infinity. A massless particle $m^2=0$, or a ``tachyon" with mass in the range $-1\leq m^2<0$, can escape to infinity. 

\item
The first term in $V_{\rm eff}(r)$ is a Coulomb-type potential that attracts the probe particle towards the event horizon at $r=r_+$.

\item
The second term in $V_{\rm eff}(r)$ acts as a {\it repulsive} Coulomb-type potential that is
sourced from the inner horizon $r=r_-$.  
\end{itemize}
 
The ``normal" situation is that the outer horizon potential is the strongest: 
$( \epsilon r_+  - j_3 r_-)^2>(\epsilon r_--j_3 r_+)^2$. Then the potential is attractive throughout spacetime. In particular, the asymptotic $\sim r^{-2}$ potential is attractive. However, the ``inverted" case $( \epsilon r_+  - j_3 r_-)^2<(\epsilon r_--j_3 r_+)^2$ is also possible. That is when the ``image" at the inner horizon is the strongest, so, at very large distance, the repulsive potential dominates. The ``marginal" case where the strengths of the two terms are identical, yields a dipole interaction. This potential is especially weak far away, because it falls off as $r^{-4}$ instead of $r^{-2}$.
The force is attractive, so it is best thought of as a limit of the ``normal" case, rather than the ``inverted". 

\medskip
\begin{figure}[h]
\centering
\includegraphics[width=10cm]{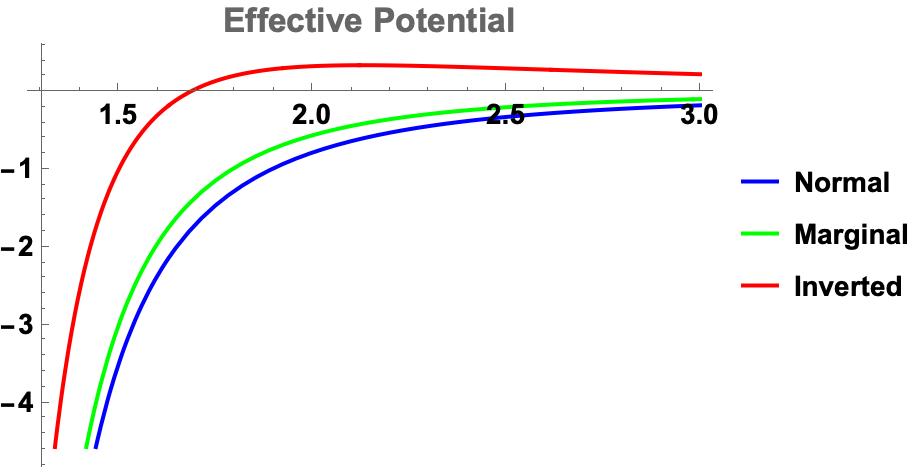}
\caption{\label{fig:effpot} The effective potential for particle motion outside the black hole. In the normal and marginal cases, the potential is attractive throughout and has overall level so particles with $m^2>0$ are confined, but those with masses in the range $-1\leq m^2\leq 0$ can escape to infinity. In the inverted case, the potential becomes repulsive far from the black hole.}
\end{figure}

It is illuminating to recast the strengths of the two potentials in terms of black hole thermodynamics. Using
\begin{equation}
\beta(1 \pm\Omega) = \frac{2\pi}{r_+\mp r_-}~,
\end{equation}
that follows from \eqref{eqn:Mrpm}
for the BTZ black holes without charges $Q_R = Q_L = 0$, we find
\begin{equation}
\left( 2\pi \cdot \frac{ \epsilon r_\pm  - j_3 r_\mp}{r^2_+-r^2_-} \right)^2 = \left( \beta(1 + \Omega) \frac{ \epsilon-j_3}{2} \pm   \beta(1 - \Omega) \frac{ \epsilon+j_3}{2}\right)^2~. 
\label{eqn:horceoeffs}
\end{equation}
The ``normal" situation is that both terms in this expression are positive. The ``inverted" potential $( \epsilon r_+  - j_3 r_-)^2<(\epsilon r_--j_3 r_+)^2$ corresponds to one of them being negative. This is precisely when one of the Boltzmann factors $\exp(  \beta(1 \pm \Omega) \frac{ \epsilon\mp j_3}{2}\big)$ corresponds to enhancement. 

Up to this point we considered only the AdS$_3$ component of the geometry, for simplicity. However, it is straigtforward to generalize the wave equation 
and take dependence on the azimuthal angles in $S^3$ into account. This is equivalent to the addition of Wilson lines $\exp\big(i q\int A_t dt)$  in the AdS$_3$ theory. 
The resulting equations are unchanged, except for the coefficients in the effective potential \eqref{eqn:Veff}. When motion on $S^3$ is included in this way, 
\eqref{eqn:horceoeffs} becomes the square of:
\begin{equation}
 \beta \Big( (1 + \Omega) \frac{ \epsilon-j_3}{2}  -  q_R\Phi_R\Big) \pm    \beta\Big((1 - \Omega) \frac{ \epsilon+j_3}{2}   -  q_L\Phi_L \Big)~.
 \label{eqn:potshift}
\end{equation}
This shows that the ``inverted" potential happens precisely when one of the
instability conditions \eqref{eqn:normalocn} is met.

The interpretation of this situation in CFT$_2$ is that, for one chirality, the population density is {\it inverted}: particles exhibit a preference for higher energy, rather than lower. This explains the terminology that distinguishes between ``normal" and ``inverted". An inverted population cannot occur spontaneously, there is no thermal equilibrium at negative temperature. However, it is possible to prepare a quantum state with inverted population and inquire how it decays to equilibrium. The effective potential \eqref{eqn:Veff} permits an interesting scenario in which particles tunnel from the near-horizon region and reach asymptotic spacetime. The inverse process is allowed as well but, as equilibrium is approached, it will be slower. Eventually, the inverted potential approaches the marginal limit, which corresponds to thermal equilibrium, with a significant component of the system banished to the asymptotic region.

\subsection{Linear Response Theory for BTZ Black Holes}
\label{subsec:response}

The standard form of the superradiant instability, discussed in section \ref{subsec:superradiant},
does not apply in AdS spacetime. 
However, linear response theory offers an analogue that relies on Green's functions, rather than a scattering  set-up. 

Consider a  bulk scalar field with ingoing boundary conditions at the black hole horizon. The asymptotic behavior of the wave function near 
the Poincar\'{e} boundary $r^{-2} \sim z=0$  takes the form: 
\begin{equation}
\psi = \psi_{(0)} \left(\frac{z}{\ell}\right)^{2-\Delta} + \psi_{(1)} \left(\frac{z}{\ell}\right)^{\Delta} ~,
\label{eqn:aswave}
\end{equation}
where the conformal weight $\Delta = 1 + \sqrt{m^2\ell^2 + 1}$. The linear response $\psi_{(1)}$ to the source $\psi_{(0)}$ gives the
retarded Green's function \cite{Herzog:2009xv,Iqbal:2011ae,Hartnoll:2016apf}: 
\begin{equation}
G_{\rm ret}(\epsilon,\vec{p})= (2\Delta -d) \frac{\psi_{(1)}}{\psi_{(0)}}~,
\label{eqn:retG}
\end{equation}
in AdS$_{d+1}$. The response function encodes the dynamics of the process. In AdS$_3$, there is only a single component of the momentum which we interpret as bulk angular momentum and refer to as $j_3$. It combines with the energy $\epsilon$ and forms the two chiral energies $\epsilon_{R/L} = \epsilon\pm j_3$. 

To compute the linear response, we solve the scalar wave equation \eqref{eqn:KGeqn}
with ingoing boundary conditions at the horizon. The result is essentially the hypergeometric function \cite{Cvetic:1997uw,Birmingham:2001pj}:
\begin{equation}
\psi(x) = \left(\frac{x-\frac{1}{2}}{x+\frac{1}{2}}\right)^{-\frac{i\epsilon}{4\pi T}}
\left( x+\frac{1}{2} \right)^{-\frac{\Delta}{2}}
~{}_2F_1 \left( \frac{\Delta}{2}- \frac{i\epsilon_L}{4\pi T_L}, \frac{\Delta}{2}- \frac{i\epsilon_R}{4\pi T_R},1 - \frac{i\epsilon}{2\pi T}, \frac{x-\frac{1}{2}}{x+\frac{1}{2}} \right)~.
\label{eqn:hypgeom}
\end{equation}
Here the dimensionless radial coordinate is $x = \frac{r^2 - \frac{1}{2}(r^2_++r^2_-)}{r^2_+-r^2_-}$. 
The wave moves towards the black hole 
because $- i (x-\frac{1}{2})\partial_x$ has negative eigenvalue near the event horizon $x\sim \frac{1}{2}$, corresponding to momentum directed towards the horizon. 
The solution with outgoing boundary conditions is the complex conjugate $\psi^*(x)$. The chiral potentials are $T^{-1}_{L,R}=\beta_{L,R} = \frac{1}{2T}(1 \pm \Omega)$ and so 
the imaginary parts of the first two arguments of ${}^2F_1$ can be recognized as $\frac{1}{2T}(1 \pm \Omega)(\epsilon\mp j_3)$. They correspond to
the two terms in the parenthesis of \eqref{eqn:horceoeffs}. Although we only write the dependence on $\epsilon_{L,R}$ explicitly, for simplicity, it is implied that motion on $S^3$ is taken into account via \eqref{eqn:potshift}.

The wave function $\psi$ \eqref{eqn:hypgeom} is written so its behavior near the outer horizon $x\sim \frac{1}{2}$ is manifest, because ${}_2F_1$ is zero when its third argument vanishes. A modular transformation of ${}_2F_1$ extracts its asymptotics as $x\to\infty$:\footnote{When $\Delta$ is integral, there are additional factors involving di-Gamma functions. They cancel the divergences of the $\Gamma$-functions when their arguments are non-positive integers.}
\begin{equation}
\psi_0(x) \sim 
x^{\frac{\Delta}{2}-1}
\frac{\Gamma(1-  \frac{i\epsilon}{2\pi T}) \Gamma(\Delta-1)}{\Gamma(\frac{\Delta}{2}-  \frac{i\epsilon}{4\pi T_L}) \Gamma(\frac{\Delta}{2}-  \frac{i\epsilon}{4\pi T_R}) }
+ x^{-\frac{\Delta}{2}} 
\frac{\Gamma(1-  \frac{i\epsilon}{2\pi T}) \Gamma(1-\Delta)}{\Gamma(1-\frac{\Delta}{2}-  \frac{i\epsilon}{4\pi T_L}) \Gamma(1-\frac{\Delta}{2}-  \frac{i\epsilon}{4\pi T_R} ) }~.
\label{eqn:psi0infty}
\end{equation}
This expression agrees with the asymptotic behavior given in \eqref{eqn:aswave} because $x\sim r^2\sim z^{-1}$. Then the retarded Green's function 
\eqref{eqn:retG} follows from the ratio of the coefficients $\psi_{(0)}, ~\psi_{(1)}$ \cite{Iqbal:2009fd,Balasubramanian:2010sc}:\footnote{For simplicity, we assume that neither of the conformal weights $\Delta_{L,R}$ are  integral or $1/2$-integral. The generalization to those cases involve di-Gamma's, as mentioned in the previous footnote. In practice, the formulae can be applied also to those cases, except that their divergent overall normalization is unphysical.}
\begin{equation}
G_{\rm ret}  (\epsilon_{L},\epsilon_R)= \frac{1}{2} \cdot
\frac{\Gamma(1 - 2\Delta_L)\Gamma(\Delta_L + \frac{\epsilon_L}{4\pi i T_L})}{(2\pi i T_L)^{1-2\Delta_L}\Gamma( 1 - \Delta_L + \frac{\epsilon_L}{4\pi i T_L})} \cdot \frac{\Gamma(1 - 2\Delta_R)\Gamma(\Delta_R + \frac{\epsilon_R}{4\pi iT_R})}{(2\pi i T_R)^{1-2\Delta_R}\Gamma( 1 - \Delta_R + \frac{\epsilon_R}{4\pi i T_R})}~. 
\label{eqn:finalret}
\end{equation}
Here we present the result for general spin $\Delta_L-\Delta_R$ even though the wave equation \eqref{eqn:KGeqn} written explicitly in this section focusses on 
a scalar with $\Delta_R=\Delta_L = \frac{1}{2}\Delta$. Introducing the lightcone coordinates $x_{L,R} = t \mp \phi$, Fourier transform with respect to $\epsilon_{L,R}$ gives the corresponding expression in real space:
\begin{equation}
G_{\rm ret}  (x_L,x_R) = - i \Theta(x_L)\Theta(x_R)\left( \frac{\pi T_L}{\sinh \pi T_L x_L}\right)^{2\Delta_L}\left( \frac{\pi T_R}{\sinh \pi T_R x_R}\right)^{2\Delta_R}~.
\label{eqn:realret}
\end{equation}
This shows that the complicated looking formula \eqref{eqn:finalret} simply expresses conformal symmetry and thermal boundary conditions. 
The $\Theta$-functions in the prefactor of \eqref{eqn:realret} give the causal structure expected for a retarded Green's function. 

The last $\Gamma$-function in the numerator of \eqref{eqn:finalret} gives poles in $G_{\rm ret}(\epsilon_R)$ when its argument 
$\Delta_{L} +  \frac{\epsilon + j_3}{4\pi i T_{L}}$ becomes a non-positive integer. Therefore, there are poles in the complex chiral energy plane when 
$$
\epsilon_L = - 4\pi i T_{L} ( \Delta_{L} + n)~, ~~n\in \mathbb{N}_0~.
$$
When $T_L$ is positive these poles are in the negative $1/2$-plane of complex $\epsilon_L$, as they should be. When the temperature is negative, they move to the upper $1/2$-plane. This indicates a sickness in the theory, that can be diagnosed as either acausality or an instabillity. Analogous considerations apply in the $R$-sector. 

For another perspective, consider the imaginary part of the response function. For bosons $\Delta_L-\Delta_R$ is integral and
\begin{equation}
{\rm Im}\,G_{\rm ret}  (\epsilon_L,\epsilon_R) = i^{2(\Delta_L-\Delta_R)} \sinh \left( \frac{\epsilon_L}{4T_{L}} + \frac{\epsilon_R}{4T_{R}} \right)
\prod_{s=R,L} \frac{(2\pi T_{s})^{2\Delta_s-1} |\Gamma(\Delta_s - \frac{i\epsilon_s}{4\pi T_{s}})|^2}{2\Gamma(2\Delta_s)\sin \pi \Delta_s} ~.
\label{eqn:ImGbos}
\end{equation}
For fermions $\Delta_L-\Delta_R$ is $1/2$-integral and
\begin{equation}
{\rm Im}\,G_{\rm ret}  (\epsilon_R,\epsilon_R) =   i^{2(\Delta_L-\Delta_R)} \cosh \left( \frac{\epsilon_L}{4T_L} + \frac{\epsilon_R}{4T_R} \right)
\prod_{s=R,L} \frac{(2\pi T_{s})^{2\Delta_s-1} |\Gamma(\Delta_s - \frac{i\epsilon_s}{4\pi T_{s}})|^2}{2\Gamma(2\Delta_s)\cos \pi \Delta_s} ~.
\label{eqn:ImGfer}
\end{equation}
The imaginary part of the retarded Green's function is proportional to the spectral function, a measure of fluctuations in the system. Importantly, it is also related to dissipation. In normal circumstances ${\rm Im}\,G_{\rm ret}>0$, corresponding to positive friction.  The formulae 
(\ref{eqn:ImGbos}-\ref{eqn:ImGfer}) have much interesting structure but, first of all, they are real and positive when $\epsilon_{R, L}>0$ and $T_{R, L}>0$. The ``inverted" potential corresponds to 
$\epsilon_{R, L}<0$ and this gives negative dissipation in the bosonic case \eqref{eqn:ImGbos}, but not for the fermion \eqref{eqn:ImGfer}. In this situation a probe particle spontaneously acquires energy from the surrounding medium. This instability is an AdS-analogue of superradiance in asymptotically flat spacetime. 

As the black hole populates the chiral primary component of the system, a dynamical equilibrium is reached when the effective temperature approaches zero. In that limit, the imaginary part of the retarded Green's function vanishes ${\rm Im}\,G_{\rm ret} (\epsilon_R,\epsilon_L) = 0$, corresponding to vanishing dissipation. If taken at face value, the resulting state is superfluid, because it has no friction. In the bulk of AdS$_3$, the vanishing of the thermodynamic potential suggests that the mode undergoes Bose-Einstein Condensation.


\section{Giant String Solutions in AdS$_3\times S^3$}
\label{sec:giantstrings}

In this section we construct giant string solutions in the AdS$_3 \times S^3$ background.
We present new classes of giant string solutions that wrap circles in both AdS$_3$ and in $S^3$.
They carry conserved charges corresponding to motion on both AdS$_3$ and $S^3$.
Generally, the strings are not supersymmetric, but we study their BPS limits.

\subsection{Preliminaries}

We first consider the AdS$_3 \times S^3$ background and the giant string action from the broader context in string theory.
This also serves to establish conventions.

A good starting point is the supergravity representation of $n_1$ D1-branes that intersect
$n_5$ D5-branes on a circle $S^1$ parametrized by the coordinate $y$.
The D5's are further wrapped around a $K3$ (or $T^4$) and viewed from a distance
where they appear smeared. Then the 10D metric becomes
\be
ds^2_{10} = (H_1 H_5)^{-\frac{1}{2}} ( - dt^2 + dy^2 ) -(H_1 H_5)^{\frac{1}{2}} (dr^2 + r^2 d\Omega^2_3)   -\left(\frac{H_5}{H_1}\right)^{\frac{1}{2}} ds^2_{K3}~, 
\ee
where the harmonic functions are $H_{1,5} = 1 + \frac{Q_{1,5}}{r^2}$.
The physical charges are related to the quantized charges through 
$Q_1 = n_1 g_s \alpha'$ and $Q_5 = n_5 g_s \alpha' $.
The approach of the harmonic functions $H_{1,5}$ to $1$ at infinity is a canonical choice
that sets the fiducial volume of the compact space to $V_{4}=(2\pi)^4\alpha'^2$.

In the near horizon region the geometry becomes AdS$_3\times S^3$:
\begin{equation}
ds^2_6 =  \ell^2( d\rho^2 + \sinh^2\rho dy^2 - \cosh^2\rho dt^2 ) 
+ \ell^2( d\theta^2 + \sin^2\theta d\xi_1^2 + \cos^2\theta d\xi_2^2)  ~,
\label{eqn:nearhAdS3}
\end{equation}
where the radii of the AdS$_3$ and the $S^3$ are
\begin{equation}
\ell = (Q_1 Q_5)^{\frac{1}{4}} = \left( n_1 n_5 (g_s \alpha')^2 \right)^{\frac{1}{4}}
= k^{\frac{1}{4}} (g_s \alpha')^{\frac{1}{2}} ~,
\label{eqn:AdS3rad}
\end{equation}
where $k \equiv n_1n_5$.

The 10D gravitational coupling $G_{10} = \frac{1}{8} (2\pi)^6 (\alpha')^4 g^2_s$ gives
$$
G_6 =  \frac{1}{8} (2\pi \alpha' g_s)^2~. 
$$
Further compactification on $S^3$ with volume $2\pi^2\ell^3$ gives the 3D gravitational coupling 
$$
G_3 =  \frac{1}{4\ell^3} (\alpha' g_s)^2~, 
$$
which, via the Brown-Henneaux formula \cite{Brown:1986nw}, gives the $SU(2)$ level: 
$$
k = \frac{c}{6} = \frac{1}{6} \cdot \frac{3\ell}{2G_3} = \frac{\ell}{4G_3} = \frac{\ell^4}{(\alpha' g_s)^2} = n_1 n_5~.
$$
This accounts for the notation $k=n_1 n_5$ that was introduced already in \eqref{eqn:AdS3rad}. 

The AdS$_3\times S^3$ geometry \eqref{eqn:nearhAdS3} is supported by
a three-form flux $*H_3$ through AdS$_3$ and the dual flux $H_3$ through $S^3$.
The corresponding two-form $C_2$ given through $H_3=dC_2$ is:
\begin{equation}
C_2 =\frac{\sqrt{k}}{2\pi\ell}  \sinh^2\rho \, dt \wedge dy + \frac{\sqrt{k}}{2\pi\ell} 
 \left(\cos^2\theta + \frac{b-1}{2} \right)\, d\xi_1 \wedge d\xi_2~.
\label{eqn:Bfields}
\end{equation}
We have introduced a gauge parameter $b$ \cite{Mandal:2007ug} that does not affect physics
and will be fixed later.
The flux is normalized such that $T \int_{S^3} *H_3 = T \int_{S^3} H_3 =k$
where $T$ is the tension of a self-dual D1/D5-string,
\begin{equation}
T = \frac{1}{2\pi\alpha' g_s} = \frac{\sqrt{k}}{2\pi \ell^2}~.
\label{eqn:Dtension}
\end{equation}
The awkward appearences of $\sqrt{k}$ effectively sets $n_1=n_5 = \sqrt{k}$,
as expected from the attractor mechanism.
The 10D dilaton $e^{-2\Phi_{10}}=\sqrt{\frac{H_5}{H_1}}$ depends on position but the
6D effective dilaton $e^{-2\Phi_{6}}$ does not. It is a modulus that keeps a record of
the division between D5- and D1-branes that the near horizon geometry \eqref{eqn:nearhAdS3}
and the scale $\ell$ given in \eqref{eqn:AdS3rad} is insensitive to \cite{Larsen:1999uk}.

The action of the probe strings we study is given by the Dirac-Born-Infeld and the Wess-Zumino terms
\be\label{stringaction0}
{\cal S} = -T \int_{\Sigma} d^2\sigma \sqrt{- \det g} + \int_\Sigma P[C_2]~,
\ee
where $T = \frac{\sqrt{k}}{2\pi \ell^2}$ is the tension of the string and
$P[C_2]$ is the pullback of the 2-form $C_2$ onto the string worldvolume $\Sigma$.

Supersymmetric instances of such probe giant string solutions were studied in the past
\cite{McGreevy:2000cw,Lunin:2002bj,Huang:2006te,Mandal:2007ug,Raju:2007uj,Prinsloo:2014dha}.
In Appendix \ref{sec:reviewGS} we review some of the progress by reinterpreting the
kappa symmetry analysis of \cite{Mandal:2007ug} in the language of complex coordinates
and holomorphic functions pioneered in \cite{Mikhailov:2000ya},
a connection that was explored in \cite{Ashok:2008fa}.
There we also provide the simplest examples of such supersymmetric solutions.
In the rest of this section, we present novel families of giant string solutions that are
generically non-supersymmetric, that may play a role in the black hole thermodynamics.

\subsection{The Equations of Motion and Solutions}

We identify the world-volume coordinates of the giant string with the AdS$_3$ coordinates $(t,y)$. Then the motion is described 
by the remaining coordinates $(\rho,\theta,\xi_1,\xi_2)$ as functions of $(t,y)$. It is governed by the Lagrangian density \eqref{stringaction0}
\bea\label{CFTaction}
\mathcal{L} &=& -T\Big[
(\cosh^2\rho - \dot\rho^2 - \dot\theta^2 - \dot\xi_1^2 \sin^2\theta - \dot\xi_2^2 \cos^2\theta)
(\sinh^2\rho + {\rho'}^2 + {\theta'}^2 + {\xi_1'}^2 \sin^2\theta  + {\xi_2'}^2 \cos^2\theta) \nn\\
&& ~~~~~~~~~~~~~~~~~~
+ (\dot\rho \rho' + \dot\theta \theta' + \dot\xi_1 \xi_1'  \sin^2 \theta+ \dot\xi_2 \xi_2' \cos^2\theta)^2 \Big]^{1/2} \nn\\
&& + T\left[\sinh^2\rho +  \left( \cos^2\theta + \frac{b-1}{2} \right) (\dot\xi_1 \xi_2' - \xi_1' \dot\xi_2) \right]~,
\eea
where dots and primes denote derivatives with respect to $t$ and $y$, respectively.
The Lagrangian density defines a 2D classical field theory,
with four fields $\rho$, $\theta$ and $\xi_{1,2}$.

From the Lagrangian density, it is a straightforward calculus to write the Euler-Lagrange equation
for each of the four fields. For example,
\bea
\frac{\partial}{\partial t} \frac{\partial \mathcal{L}}{\partial \dot\rho}
+ \frac{\partial}{\partial y} \frac{\partial \mathcal{L}}{\partial \rho'}
- \frac{\partial \mathcal{L}}{\partial \rho} &=& 0~.
\eea
It seems impossible to solve the system of nonlinear differential equations generally, so we restrict to
solutions where both of the $S^3$ azimuthal angles $(\xi_1,\xi_2)$ are linear functions of $t$ and $y$
and both of $\rho$ and $\theta$ are constants:  
\bea\label{linearxi}
\begin{cases}
\xi_1 = a_1 t + b_1 y~, & \\
\xi_2 = a_2 t + b_2 y~, &
\end{cases}
\qquad \dot\rho = \rho' = \dot\theta = \theta' = 0~.
\eea

This class of giant string solutions are in general neither the giant
(pointlike in AdS$_3$ and wrapping $S^1 \subset S^3$)
nor the dual-giant (pointlike in $S^3$ and wrapping $S^1 \subset {\rm AdS}_3$).
At each time $t$, the $S^3$ angles $(\xi_1,\xi_2)$ are linear functions of the azimuthal angle $y$ on AdS$_3$. 
This corresponds to a circle in a diagonal direction inside both the AdS$_3$ and the $S^3$ geometries.
This is a novel feature of our class of giant string solutions that may hint towards an analogous class of D3-brane solutions in AdS$_5 \times S^5$.

After the restriction \eqref{linearxi}, the $\xi_{1,2}$ equations of motion are solved automatically. 
The $\rho$ equation of motion becomes
an algebraic equation,
\begin{small}
\bea\label{eomrho}
\hspace{-.9cm} 
0 = \cosh\rho \sinh\rho \cdot \left[
\cosh^2\rho + \sinh^2\rho - (a_1^2 - b_1^2)\sin^2\theta - (a_2^2 - b_2^2) \cos^2\theta - 2\cdot \frac{\sqrt{-\det g}}{\ell}
\right]~,
\eea
\end{small}
where
\begin{small}
\bea\label{sqrtg}
\!\!\!\!\!\! \frac{\! \sqrt{\!-\! \det g}}{\ell} \!\!\! &=& \!\!\! \sqrt{(\cosh^2\rho - a_1^2 \sin^2\theta - a_2^2 \cos^2\theta)
(\sinh^2\rho + b_1^2 \sin^2\theta  + b_2^2 \cos^2\theta) + (a_1b_1 \sin^2 \theta + a_2b_2 \cos^2\theta)^2}~. \\
 \!\!\! &=&  \!\!\! \sqrt{\! \cosh^2 \! \rho \sinh^2 \! \rho\!-\! (a_1^2 \sin^2\!\theta \!+\! a_2^2 \cos^2\!\theta)\sinh^2 \! \rho
\!+\! (b_1^2 \sin^2\theta \!+\! b_2^2 \cos^2\theta) \cosh^2\! \rho \!-\! (a_1b_2 \!-\! a_2b_1)^2 \cos^2\!\theta\sin^2\!\theta}~. \nn
\eea
\end{small}
There is a trivial solution to \eqref{eomrho} given by $\rho=0$.
Otherwise, for the square bracket to vanish, one obtains
\bea\label{eomrho2}
(a_1 \pm b_1)^2 \sin^2\theta + (a_2 \pm b_2)^2 \cos^2\theta &=& 1~.
\eea
The sign ambiguity arises from squaring \eqref{sqrtg},
so one must choose the sign in \eqref{eomrho2} such that $- \det g > 0$.
The latter imposes the extra condition that\footnote{
We impose a strict inequality, because equality yields solutions with vanishing worldvolume.
The analogous reasoning will apply to \eqref{eomtheta3}.}
\bea\label{eomrho3}
\cosh^2\rho + \sinh^2\rho - (a_1^2 - b_1^2)\sin^2\theta - (a_2^2 - b_2^2) \cos^2\theta &>& 0~.
\eea
For each branch, \eqref{eomrho2} itself has two branches of solutions. If the constants are such that
$(a_1 \pm b_1)^2 = (a_2 \pm b_2)^2 = 1$, then \eqref{eomrho2} is satisfied identically for all polar angles $\theta$,
but otherwise it is an equation for $\theta$.
To summarize, the $\rho$ equation of motion \eqref{eomrho} has three branches of solution:
\bea\label{eomrhosol}
\text{branch $\rho$A:} && \rho = 0~, \nn\\
\text{branch $\rho$B:} && (a_1 \pm b_1)^2 = (a_2 \pm b_2)^2 = 1~,\\
\text{branch $\rho$C:} && 
\cos^2\theta = \frac{(a_1 \pm b_1)^2 - 1}{(a_1 \pm b_1)^2 - (a_2 \pm b_2)^2} \quad\leftrightarrow\quad
\sin^2\theta = \frac{1-(a_2 \pm b_2)^2}{(a_1 \pm b_1)^2 - (a_2 \pm b_2)^2} ~. \nn
\eea
Note that the $\rho$C branch assumes $\rho$B is not satisfied, or else denominator in the formula vanishes.

The $\theta$ equation of motion 
\bea\label{eomtheta}
0 = \sin\theta \cos\theta \cdot && \!\!\! \Big[
(a_1^2-a_2^2) \sinh^2\rho - (b_1^2-b_2^2) \cosh^2\rho + (a_1b_2 - a_2b_1)^2 (\cos^2\theta - \sin^2\theta) 
~ \nn\\
 && - 2(a_1b_2-a_2b_1) \cdot \frac{\sqrt{-\det g}}{l}
\Big] ~,
\eea
yields analogous branches of solutions. It is solved  trivially by $\theta = 0$ or $\frac{\pi}{2}$.
Otherwise, one must solve
\bea\label{eomtheta2}
(b_1 \pm b_2)^2 \cosh^2 \rho - (a_1 \pm a_2)^2 \sinh^2 \rho &=& (a_1b_2 - a_2b_1)^2~. 
\eea
The signs much be chosen such that $- \det g > 0$, which is equivalent to: 
\bea\label{eomtheta3}
\frac{(a_1^2-a_2^2) \sinh^2\rho - (b_1^2-b_2^2) \cosh^2\rho}{a_1b_2 - a_2b_1}
+ (a_1b_2 - a_2b_1) (\cos^2\theta - \sin^2\theta) &>& 0~.
\eea
The equation \eqref{eomtheta2} is satisfied identically for all $\rho$ if
$(a_1 \pm a_2)^2 = (b_1 \pm b_2)^2 = (a_1b_2 - a_2b_1)^2$,
and otherwise it is an equation for $\rho$.
To summarize, the $\theta$ equation of motion \eqref{eomtheta} has three branches of solution:
\bea\label{eomthetasol}
\text{branch $\theta$A:} && \theta = 0 ~~\text{or}~~ \frac{\pi}{2} ~, \nn\\
\text{branch $\theta$B:} && (a_1 \pm a_2)^2 = (b_1 \pm b_2)^2 = (a_1b_2 - a_2b_1)^2~, \\
\text{branch $\theta$C:} &&
\cosh^2\rho = \frac{(a_1 \pm a_2)^2 - (a_1b_2 - a_2b_1)^2}{(a_1 \pm a_2)^2 - (b_1 \pm b_2)^2} ~~\leftrightarrow~~
\sinh^2\rho = \frac{(b_1 \pm b_2)^2 - (a_1b_2 - a_2b_1)^2}{(a_1 \pm a_2)^2 - (b_1 \pm b_2)^2}~. \nn
\eea
Again  the $\theta$C branch assumes $\theta$B is not satisfied, or else denominator in the formula vanishes.

Any combination of one of the branches of \eqref{eomrhosol} and one of \eqref{eomthetasol}
defines a classical giant string solution.
Each branch A or C imposes one condition on the six continuous parameters
$(a_1,a_2,b_1,b_2,\rho,\theta)$, and each branch B imposes two conditions.
As branches $\rho$C and $\theta$C determine the value of
$\cos^2\theta$ and $\cosh^2\rho$, respectively,
these yield valid physical solutions only if $0 \leq \cos^2\theta \leq 1$ and/or $1 \leq \cosh^2\rho$.
Parameters for branches other than $\rho$A and $\theta$A are also bounded by the inequalities
\eqref{eomrho3} and \eqref{eomtheta3}. 

A generic family of solutions to both equations of motion are via branches $\rho$C and $\theta$C. In this case the equations of motion 
determine $\rho$ and $\theta$ in terms of the four parameters $(a_1,a_2,b_1,b_2)$.
We refer to this four-parameter ($\rho$C, $\theta$C) family of string solutions as Type I solutions,
and it will be discussed in detail in section \ref{sec:CCsoln}.

Although branches B each imposes two constraints, it is possible to obtain 
a four-parameter family of solutions also via branches $\rho$B and $\theta$B,
because the branch $\rho$B may imply $\theta$B. Let us elaborate on this.
Consider the branch $\rho$B for solution of the $\rho$ equation of motion.
Then the parameters $(a_1,a_2,b_1,b_2)$ satisfy
\bea\label{BB4constraints}
s_1 (a_1 - s_3 b_1) = a_2 - s_3 b_2 = -s_1s_2s_3~,
\eea
for some choice of signs $s_1,s_2,s_3 = \pm 1$. Elementary reorganizations of these equations give:
\bea
b_1-s_1b_2 = s_3 (a_1-s_1a_2)~, \qquad
a_1b_2 - a_2b_1 = s_1s_2 (a_1-s_1a_2)~. 
\label{BB4constraintsV2}
\eea
These give the conditions for $\theta$B with the sign $\pm = -s_1$. 

With the sign choices in \eqref{BB4constraints}, the bound \eqref{eomrho3} becomes:
\bea\label{eomrho3rep}
0 &<& \cosh^2\rho + \sinh^2\rho - (a_1^2 - b_1^2)\sin^2\theta - (a_2^2 - b_2^2) \cos^2\theta \nn\\
&=& \cosh^2\rho + \sinh^2\rho + s_2s_3(a_1 + s_3b_1)\sin^2\theta +s_1s_2s_3 (a_2 + s_3b_2)\cos^2\theta \nn\\
&=& 2 \cosh^2\rho + 2 s_2s_3 (a_1 \sin^2\theta + s_1a_2 \cos^2\theta)~.
\eea
Similarly, with the sign choices in \eqref{BB4constraints}, the bound \eqref{eomtheta3} imposed on the $\theta$B branch becomes:
\bea\label{eomtheta3rep}
0 &<& \frac{(a_1-s_1a_2)(a_1+s_1a_2) \sinh^2\rho - (b_1-s_1b_2)(b_1+s_1b_2) \cosh^2\rho
+ (a_1b_2 - a_2b_1)^2 (\cos^2\theta - \sin^2\theta)}{a_1b_2 - a_2b_1} \nn\\
&=& \frac{(a_1+s_1a_2) \sinh^2\rho - s_3(b_1+s_1b_2) \cosh^2\rho
+ (a_1 - s_1a_2) (\cos^2\theta - \sin^2\theta)}{s_1s_2} \nn\\
&=& - 2s_1s_3 \Big(\cosh^2\rho + s_2s_3 (a_1 \sin^2\theta + s_1 a_2 \cos^2\theta)\Big)~.
\eea
Consistency between \eqref{eomrho3rep} and \eqref{eomtheta3rep} requires that the signs satisfy $s_1s_3=-1$.
With this sign choice understood, we have shown that the $\rho$B branch of solutions
implies the $\theta$B branch. Moreover, the computations also hold in the opposite direction: $\theta$B implies $\rho$B.
Further, it follows that combinations of ($\rho$B, $\theta$C) or ($\rho$C, $\theta$B) branches are inconsistent.

We have established that the two constraints \eqref{BB4constraints}, with signs that satisfy $s_1s_3=-1$, are
enough to solve the equations imposed by both of the branches ($\rho$B, $\theta$B). This
results in another four parameter family of string solutions parametrized by two of
$(a_1,a_2,b_1,b_2)$ that we choose to be $(a_1,a_2)$, $\rho$ and $\theta$.
We refer to this family as Type II solutions.
It is as generic as, but distinct from, the Type I family.
It will be discussed in detail in section \ref{sec:BBsoln}.

The ($\rho$A, $\theta$A) branch reduces to the giants gravitons that were found long time ago and reviewed in the Appendix. 
We have found no interesting mixtures between the $\rho$A and $\theta$A and those of other types. 

\subsection{Symmetries and Conserved Charges}\label{sec:symcharge}

All our string solutions respect the four obvious continuous symmetries: translation of the AdS$_3$ coordinates $(t,y)$ and translations of the azimuthal angles $(\xi_1, \xi_2)$.
In this subsection we compute the corresponding conserved charges $(E, J_3)$ and $(Q_1,Q_2)$. 
These four charges will offer a parametrization of our solutions that is complementary to $(a_1, a_2, b_1, b_2)$.  

For the constant-$(\rho,\theta)$ and linear-$\xi$ \eqref{linearxi} solutions that we are interested in, charge densities will 
not depend on $y$. Therefore, the integral over $y$ that converts from charge densities to the charges itself is just a multiplication by $2\pi$.
All charges will also be proportional to the string tension. Combining these features, a universal rescaling of the dimensionful parameters simplifes all prefactors:
\be\label{rescalesqrtk}
\ell T \to \frac{1}{2\pi} \qquad \Leftrightarrow \qquad \frac{\sqrt{k}}{\ell} \to 1~.
\ee

With these normalizations, the momentum densities conjugate to $(\xi_1,\xi_2)$ are given through 
\bea\label{CFTj}
2\pi  \frac{\partial  \mathcal{L}}{\partial \dot\xi_1} & = &
 \frac{\dot\xi_1 \sin^2\theta \sinh^2\rho
+ \xi_2' \sin^2\theta \cos^2\theta (\dot\xi_1 \xi_2' - \xi_1' \dot\xi_2)}{\sqrt{-\det g}/\ell}
+ \left( \cos^2\theta + \frac{b-1}{2} \right) \xi_2' ~, \nn\\
2\pi  \frac{\partial  \mathcal{L}}{\partial \dot\xi_2} & = &
 \frac{\dot\xi_2 \cos^2\theta \sinh^2\rho
- \xi_1' \sin^2\theta \cos^2\theta (\dot\xi_1 \xi_2' - \xi_1' \dot\xi_2)}{\sqrt{-\det g}/\ell}
- \left( \cos^2\theta + \frac{b-1}{2} \right) \xi_1' ~. \qquad
\eea
These expressions depend on the gauge parameter $b$. However, any $b$ is physically equivalent, it just changes the conserved charges by a constant. We adapt the gauge choice of \cite{Mandal:2007ug},
which is to choose $b=1$ for $q_1$ and $b=-1$ for $q_2$. Thus,
\bea\label{CFTJ}
q_1 &=& \frac{a_1 \sin^2\theta \sinh^2\rho
+ b_2 \sin^2\theta \cos^2\theta (a_1 b_2 - b_1 a_2)}{\sqrt{-\det g}/\ell}
+ b_2 \cos^2\theta ~, \nn\\
q_2 &=& \frac{a_2 \cos^2\theta \sinh^2\rho
- b_1 \sin^2\theta \cos^2\theta (a_1 b_2 - b_1 a_2)}{\sqrt{-\det g}/\ell}
+ b_1 \sin^2\theta ~.
\eea

The other two continuous symmetries are spacetime translations of $t$ and $y$.
The corresponding conserved currents combine to the canonical energy momentum tensor
$${T^\mu}_\nu = \sum_{i=1,2} \frac{\partial \mathcal{L}}{\partial (\partial_\mu \xi_i)} \partial_\nu \xi_i
- \mathcal{L} {\delta^\mu}_\nu~.$$
The time components $(\mu=t)$ for the two currents $(\nu=t,y)$ give the Hamiltonian and the momentum density:
\bea
\mathcal{H} &\equiv& \dot\xi_1\frac{\partial  \mathcal{L}}{\partial \dot\xi_1} + \dot\xi_2 \frac{\partial  \mathcal{L}}{\partial \dot\xi_2}  - \mathcal{L} ~,\nn\\
\mathcal{J}_3 &\equiv& \xi'_1\frac{\partial  \mathcal{L}}{\partial\dot\xi_1} + \xi'_2 \frac{\partial  \mathcal{L}}{\partial \dot\xi_2}~.
\eea
The notation $\mathcal{J}_3$ reminds us that  momentum within AdS$_3$ is an angular momentum. When evaluated on the class of solutions  \eqref{linearxi}, 
the conserved charges become
\bea\label{CFTEP}
\ep &=& \frac{\cosh^2\rho \left( \sinh^2\rho + b_1^2 \sin^2\theta  + b_2^2 \cos^2\theta \right)}{\sqrt{-\det g}/\ell}
- \sinh^2\rho~, \nn\\
j_3 &=& \frac{\sinh^2\rho \left( a_1b_1 \sin^2\theta  + a_2b_2 \cos^2\theta \right)}{\sqrt{-\det g}/\ell}~.
\eea
Both of these are independent of the gauge parameter $b$. The formulae for conserved charges \eqref{CFTJ} and \eqref{CFTEP} will be essential for computations in the remainder of the section. For complete detail, they must be augmented with the expression for $\sqrt{-\det g}$ given in \eqref{sqrtg}. 
%
%
Also we will eventually restore the overall factors simplified in \eqref{rescalesqrtk}. They show that the actual charges scale as $\sqrt{k}$, and that they appear dimensionless only because they have been scaled by $\ell$.

The system we study has two discrete $\mathbb{Z}_2$-symmetries.
The Lagrangian \eqref{CFTaction} is unchanged when simultaneously flipping the signs of $y$ (and so the prime variables)
and $\xi_1$. It is also invariant when flipping the signs of both $t$ (and so the dotted variables) and $\xi_2$.
Either way, it is equivalent to flipping the signs of $(a_1,b_2)$ in the linear combinations \eqref{linearxi}.
Similarly, the Lagrangian is unchanged when simultaneously flipping the signs of $y$ and $\xi_2$, or flipping the signs of $t$ and $\xi_1$.
These operations both correspond to flipping the signs of $(a_2,b_1)$ in \eqref{linearxi}.
Taken together, the two discrete symmetries map any string solution to three others, unless there is an exceptional degeneracy.

The conserved quantities of the four solutions related by $\mathbb{Z}_2 \times \mathbb{Z}_2$ symmetry are also closely related,
as apparent from from \eqref{CFTJ}, \eqref{CFTEP} and \eqref{sqrtg}.
By flipping the signs of $(a_1,b_2)$, $(j_3,q_1)$ are flipped,
and by flipping the signs of $(a_2,b_1)$, $(j_3,q_2)$ are flipped.
The combination of the two operations flip the signs of $(q_1,q_2)$. In terms of the chiral charges defined as
$$
e_R \equiv \ep+j_3~, ~e_L \equiv \ep-j_3~, ~q_R \equiv q_1+q_2~, ~q_L \equiv q_1-q_2~,
$$
the $\mathbb{Z}_2$ operation of flipping $(a_2,b_1)$ is equivalent to interchanging $L \leftrightarrow R$,
while the $\mathbb{Z}_2$ operation of flipping all of $(a_1,a_2,b_1,b_2)$ is equivalent to
flipping the signs of $q_{L,R}$. This is a useful choice of basis for the $\mathbb{Z}_2 \times \mathbb{Z}_2$ discrete symmetry.
In the following, we will often study only one of the four equivalent cases, without loss of generality.

\subsection{Type I ($\rho$C, $\theta$C) Solutions}
\label{sec:CCsoln}

In this subsection we study in detail the four-parameter family of giant string solutions
that are obtained by solving the $\rho$ and $\theta$ equations of motion via the branches $\rho$C and $\theta$C.
This family is parametrized by four real parameters $(a_1,a_2,b_1,b_2)$ that, according to (\ref{eomrhosol}, \ref{eomthetasol}) determine the constant values
of $\rho$ and $\theta$ as:
\bea\label{CCrhotheta}
\cos^2\theta &=& \frac{(a_1 \pm b_1)^2 - 1}{(a_1 \pm b_1)^2 - (a_2 \pm b_2)^2}~, \nn\\
\cosh^2\rho &=& \frac{(a_1 \pm a_2)^2 - (a_1b_2 - a_2b_1)^2}{(a_1 \pm a_2)^2 - (b_1 \pm b_2)^2}~. 
\eea
The parameters $(a_1,a_2,b_1,b_2)$ must be such that $0 \leq \cos^2\theta \leq 1$ and $1\leq \cosh^2\rho$.
The signs $\pm$ are identical within each line of \eqref{CCrhotheta}, but they may differ between the two equations, for now. 
Moreover, according to (\ref{eomrho3}, \ref{eomtheta3}) they must be such that:
\bea\label{CCvalidate}
\cosh^2\rho + \sinh^2\rho - (a_1^2 - b_1^2)\sin^2\theta - (a_2^2 - b_2^2) \cos^2\theta &>& 0~, \nn\\
\frac{(a_1^2-a_2^2) \sinh^2\rho - (b_1^2-b_2^2) \cosh^2\rho}{a_1b_2 - a_2b_1}
+ (a_1b_2 - a_2b_1) (\cos^2\theta - \sin^2\theta) &>& 0~.
\eea
For a valid set of parameters, the charges of the solution can be computed
from (\ref{CFTJ}, \ref{CFTEP}).

In order to satisfy the inequalities \eqref{CCvalidate}, one must choose $+$ ($-$) for
the signs in both lines of \eqref{CCrhotheta} if
\be\label{CCdetsign}
\frac{a_1b_1 (1-a_2^2-b_2^2) - a_2b_2 (1-a_1^2-b_1^2)}{(a_1+a_2-b_1-b_2)(a_1-a_2+b_1-b_2)}~,
\ee
is positive (negative). Either of the two discrete $\mathbb{Z}_2$ symmetries that flips $(a_2,b_1) \to (-a_2,-b_1)$ or
or $(a_1,b_2) \to (-a_1, -b_2)$, respectively, flips the sign of the sign decider \eqref{CCdetsign}.
Therefore, we can restrict to the subset of parameters $(a_1,a_2,b_1,b_2)$ for which \eqref{CCdetsign} is positive, and take $+$ signs throughout \eqref{CCrhotheta}. There is no loss of generality because all solutions with negative \eqref{CCdetsign} and $-$ signs throughout \eqref{CCrhotheta}
can be recovered easily using the discrete symmetry.
With this sign choice, we rewrite the physical range for $\rho$ and $\theta$ as
\bea\label{CCcond}
0 \leq \cos^2\theta = \frac{(a_1 + b_1)^2 - 1}{(a_1 + b_1)^2 - (a_2 + b_2)^2} \leq 1~, \nn\\
1 \leq \cosh^2\rho = \frac{(a_1 + a_2)^2 - (a_1b_2 - a_2b_1)^2}{(a_1 + a_2)^2 - (b_1 + b_2)^2}~.
\eea
In summary, the physical region for the parameters $(a_1,a_2,b_1,b_2)$ is where \eqref{CCdetsign} is positive and \eqref{CCcond} are satisfied. 

Substitution of the expressions \eqref{CCcond} for $\rho$ and $\theta$ in the formulae (\ref{CFTJ},\ref{CFTEP}) for the four conserved charges gives: 
\bea\label{CCcharges}
\frac{e_R}{2} ~=~ \frac{\ep+j_3}{2}
&=& \frac{b_1+b_2}{a_1+a_2+b_1+b_2}~, \nn\\
\frac{e_L}{2} ~=~ \frac{\ep-j_3}{2}
&=& \frac{(a_1b_2 - a_2b_1)^2 + (a_1+a_2)(b_1+b_2)}{(a_1+a_2+b_1+b_2)^2}~, \nn\\
\frac{q_R}{2} ~=~\frac{q_1+q_2}{2}
&=& \frac{(b_1+b_2)(1+(a_1+b_1)(a_2+b_2))}{(a_1+a_2+b_1+b_2)^2}~, \nn\\
\frac{q_L}{2} ~=~ \frac{q_1-q_2}{2}
&=& \frac{a_1b_2 - a_2b_1}{a_1+a_2+b_1+b_2}~.
\eea
These formulae automatically satisfy the BPS bounds
\bea\label{CCBPSbounds}
e_R \geq |q_R|~, \qquad e_L \geq |q_L|~,
\eea
provided that $(a_1,a_2,b_1,b_2)$ are in the physical region.

Although the four charges are parametrized by four parameters, the map is not one-to-one. That
is because the charges \eqref{CCcharges} always satisfy one relation:
\bea\label{CCchargerel}
\frac{e_L}{2} - \left(\frac{q_L}{2} \right)^2
&=& \frac{e_R}{2} - \left(\frac{e_R}{2} \right)^2~.
\eea
Therefore, the four-parameter Type I family of giant string solutions only cover
a 3d subspace of the 4d charge space $(e_R,e_L,q_R,q_L)$. The constant radial coordinate $\rho$ is also a function of the charges,
\be
2\cosh^2\rho = \frac{2- e_R-e_L}{1-e_R}
\quad \Leftrightarrow \quad
2 \sinh^2\rho = \frac{e_R-e_L}{1-e_R}~,
\label{eqn:rhoIEREL}
\ee
but $\theta$ is not. Thus, the Type I giant string solutions can be parametrized by
four \emph{physical} parameters, namely 3 of the 4 charges and $\theta$.

\begin{figure}
\begin{center}
\includegraphics[width=0.48\textwidth]{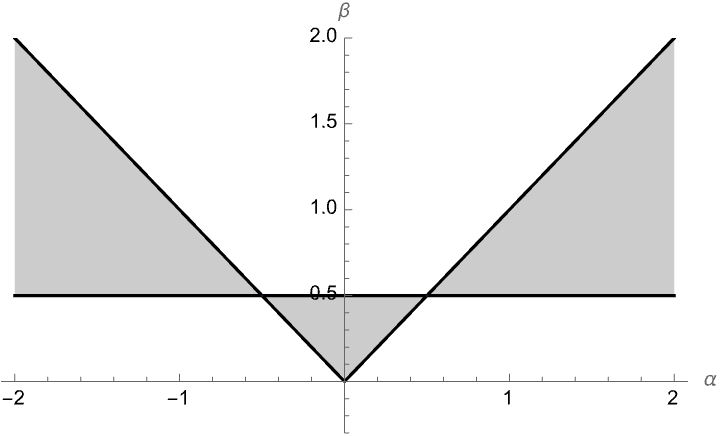}
\includegraphics[width=0.48\textwidth]{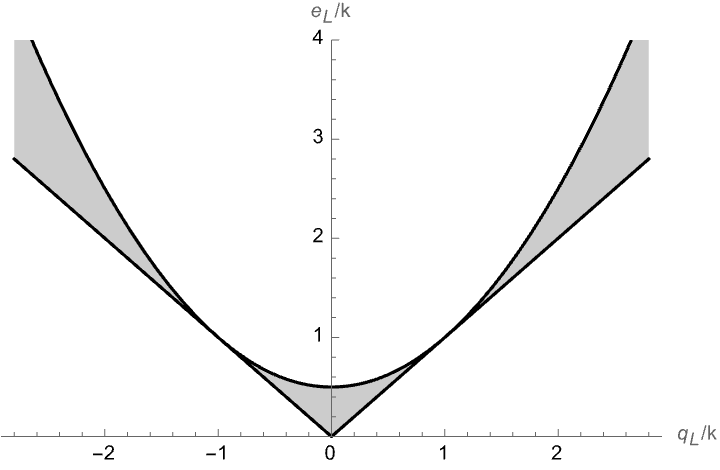}
\caption{\label{Fig:abgallowed} The physical region for the parameters $(\alpha,\beta)$ of
Type I solutions, and for their charges $(q_L,e_L)$.
Given $(q_L,e_L)$ within the shaded region, $e_R$ is determined by \eqref{CCchargerel},
and the allowed range of $q_R$ is simply $|q_R| \leq e_R$.}
\end{center}
\end{figure}

As we will be mostly interested in the charges of the giant string solution,
it is useful to introduce the following three parameters,
\bea\label{CCabg}
\hspace{-.8cm} \alpha \equiv \frac{a_1b_2 - a_2b_1}{a_1+a_2+b_1+b_2}~, \quad
\beta \equiv \frac{b_1 + b_2}{a_1+a_2+b_1+b_2}~, \quad
\gamma \equiv \frac{1+(a_1+b_1)(a_2+b_2)}{a_1+a_2+b_1+b_2}~.
\eea
In terms of these, the four charges (of which only three are independent) are written as
\bea\label{CCchargesabg}
\frac{e_R}{2} = \beta~, \quad
\frac{e_L}{2} = \alpha^2 + \beta (1-\beta) ~, \quad
\frac{q_R}{2} = \beta \gamma ~, \quad
\frac{q_L}{2} = \alpha~.
\eea
The three numbers $(\alpha,\beta,\gamma)$ parametrize all four charges of the giant string solution, while ensuring
that the constraint \eqref{CCchargerel}) is satisfied. To single out a member of the Type I solutions, 
one further needs to specify $\theta$. The advantage of using the parameters $(\alpha,\beta,\gamma)$ is that the 
physical region, where \eqref{CCdetsign} is positive and the bounds \eqref{CCcond} are satisfied,
can be expressed quite simply and linearly. The necessary and sufficient condition for $(\alpha,\beta,\gamma)$ to parametrize
a physical Type I solution, can be summarized as
\bea\label{condabg}
\begin{cases}
\beta < 0 & \text{: not allowed}~, \\
0 \leq \beta < \frac12 & \to |\alpha| \leq \beta~, \\
\frac12 < \beta & \to |\alpha| \geq \beta~,
\end{cases}
\qquad \text{and} \quad |\gamma| \leq 1~.
\eea
The allowed range for $(\alpha,\beta)$ is closely related to the physical condition that
\be\label{CCrho}
\sinh^2\rho = \frac{e_R-e_L}{2(1-e_R)}
= \frac{\beta^2 -\alpha^2}{1-2\beta} \geq 0~.
\ee

The expressions \eqref{CCchargesabg} of charges as simple functions of $(\alpha,\beta,\gamma)$
are useful for understanding the physical region.
For example, the BPS bounds \eqref{CCBPSbounds} are easy to recover from 
\eqref{CCchargesabg} and \eqref{condabg}, but more elaborate manipulations are needed in the parametrization
\eqref{CCcharges}. The allowed range \eqref{condabg} also shows: 
\be
\frac{e_L}{2} - \left(\frac{q_L}{2} \right)^2 = \beta(1-\beta) \leq \frac14~.
\ee
In Figure \ref{Fig:abgallowed}, the left plot gives the allowed range \eqref{condabg}
of the parameters $(\alpha,\beta)$ and the right plot shows the physical region for the charges
$(q_L,e_L)$.

The physics of the dual CFT$_2$ should be the same in the left and right copies of the algebra, 
but our formulae have an apparent asymmetry: the constraint \eqref{CCchargerel} and the 
parametrization \eqref{CCchargesabg} are not invariant
under the interchange $(e_L,q_L) \leftrightarrow (e_R,q_R)$. 
That is possible because the discrete $\mathbb{Z}_2$ symmetry that flips the sign of $(a_2,b_1)$ interchanges the
left and the right charges, and it also flips the sign of \eqref{CCdetsign}.
Therefore, for any solution described in this subsection, there is another solution
obtained by the $\mathbb{Z}_2$ operation, whose charges differ only by $(e_L,q_L) \leftrightarrow (e_R,q_R)$.
The combination of the two families cover a symmetric 3d subspace of the 4d charge space.

\subsubsection{The Supersymmetric Limits}

The charges of the Type I giant string solutions, expressed simply in terms of
$(\alpha,\beta,\gamma)$, manifestly satisfy the BPS bounds \eqref{CCBPSbounds}.
We are interested in their saturation, the supersymmetric limits of the giants.

According to the parametrizations \eqref{CCchargesabg}, saturation of $e_R \geq |q_R|$ is achieved by $\gamma = \pm 1$.
Since $q_R$ is the only charge that depends on $\gamma$, this supersymmetric limit does not involve other charges.
Equivalently, because the charge constraint \eqref{CCchargerel} does not involve $q_R$, we can freely tune
$q_R$ to $\pm e_R$, whatever the values of the other three charges. It follows that the $\frac14$-BPS giant string solutions with $e_R = |q_R|$
cover a 2d subspace of the 4d charge space. It can be parametrized by $(\alpha, \beta)$:
\bea\label{CCSUSY1}
\frac{e_R}{2} = \pm \frac{q_R}{2} = \beta~, \quad
\frac{e_L}{2} = \alpha^2 + \beta (1-\beta) ~, \quad
\frac{q_L}{2} = \alpha~.
\eea
The physical range for $(\alpha,\beta)$ is as described in \eqref{condabg} and in Figure \ref{Fig:abgallowed}.

Recall that, for the Type I solutions, $\theta$ is the ``extra" parameter that is independent of the charges. 
However, from the definition \eqref{CCabg} of $\gamma$, we can show that $\theta$ given in \eqref{CCcond} is fixed as well:
\be
\gamma = \pm 1 \quad \Leftrightarrow \quad (a_1+b_1 \pm 1)(a_2+b_2 \pm 1) = 0
\quad \Leftrightarrow \quad \theta = 0 ~\text{or}~ \frac{\pi}{2} ~,
\ee
Therefore, the $\frac14$-BPS condition $e_R = \pm q_R$ imposes
\emph{two} constraints on the four-parameter family of Type I solutions. 
The BPS subclass is a two-parameter family that we can parametrize as \eqref{CCSUSY1}.

We now turn to saturation of $e_L \geq |q_L|$.
If $e_L = \pm q_L$, \eqref{CCchargesabg} implies
\be
\beta (1-\beta) = \pm \alpha (1 \mp \alpha) \quad\Rightarrow\quad
\begin{cases}
\beta = \pm \alpha~, \\
\beta \pm \alpha = 1~.
\end{cases}
\ee
However, the physical region \eqref{condabg} generally disallows $\beta = 1\mp \alpha $, with the only 
exception at $\beta = \pm \alpha\pm \frac{1}{2}$. This conclusion also follows from Figure \ref{Fig:abgallowed}.
It follows that $\beta = \pm \alpha \geq 0$ in all cases.
This family of $\frac14$-BPS giant string solutions also covers a 2d subspace of the 4d charge space, here 
parametrized by $\alpha$ and $\gamma$:
\bea\label{CCSUSY2}
\frac{e_R}{2} = |\alpha| ~, \quad
\frac{e_L}{2} = \text{sign}\,(\alpha) \cdot \frac{q_L}{2} = |\alpha| ~, \quad
\frac{q_R}{2} = |\alpha| \cdot \gamma ~,
\eea
where $-1 \leq \gamma \leq 1$.
Notably, $e_R = e_L$ in this class of $\frac14$-BPS solutions.
This further implies $\rho = 0$ due to \eqref{CCrho}, whereas $\theta$ remains unfixed.

The $\frac12$-BPS Type I giant string is an intersection of the $\frac14$-BPS 
families \eqref{CCSUSY1} and \eqref{CCSUSY2}.
The charges satisfy
\bea\label{CCSUSY3}
e_R = e_L = |q_R| = |q_L|~,
\eea
and both $\rho = 0$ and $\theta = 0 ~\text{or}~ \frac{\pi}{2}$ are fixed.
This is a one-parameter family of solutions. A conclusion that is relevant for the rest of this paper is that
all $\frac12$-BPS giant string solution of Type I has $e_R = e_L$, equivalently $j_3= 0$.

\subsection{Type II ($\rho$B, $\theta$B) Solutions}
\label{sec:BBsoln}

In this subsection we study in detail the four-parameter family of giant string solutions
that are obtained by solving the $\rho$ and $\theta$ equations of motion
via the branches $\rho$B and $\theta$B defined in \eqref{eomrhosol} and \eqref{eomthetasol}. 
This Type II family is as generic as the Type I studied in the previous subsection
in the sense that it is a four-parameter family and that it covers
a 3d subspace of the 4d charge space,
but it is distinct from the Type I solutions.

For the ($\rho$B, $\theta$B) branch, the $\rho$ and $\theta$ equations of motion are solved 
precisely when \eqref{BB4constraints} is satisfied with $s_1s_3=-1$, {\it wiz.}
\bea\label{BBcond1}
s_1 (a_1 + s_1 b_1) = a_2 +s_1 b_2 = s_2~. 
\eea
The signs $s_1$ and $s_2$ are independenly $\pm 1$, but we can exploit the two
$\mathbb{Z}_2$ symmetries to fix both signs $s_{1,2}$ without losing generality.
The $\mathbb{Z}_2$ operation that flips the sign of $(a_1,b_2)$ corresponds in \eqref{BBcond1} to flipping $s_1$,
and the complementary $\mathbb{Z}_2$ that flips $(a_2,b_1)$ corresponds to flipping both $s_{1,2}$.
Therefore, without loss of generality, we can choose $s_1=s_2=+1$ (which implies $s_3=-1$) and so
\be\label{BBbina}
b_1 = 1-a_1~, \qquad b_2 = 1-a_2~. 
\ee
The $(a_1,a_2)$ remain as two independent parameters.

For the Type I ($\rho$C, $\theta$C) solutions presented in the previous subsection, the parameters
$(a_1,a_2,b_1,b_2)$ determine constant values of the coordinates $\rho$ and $\theta$.
In contrast, for the Type II ($\rho$B, $\theta$B) solutions studied in this subsection, $\rho \geq 0$ and $0 \leq \theta \leq \frac{\pi}{2}$ are two free parameters
in addition to $(a_1, a_2)$. In this case the physical conditions $0 \leq \cos^2\theta \leq 1$ and $1\leq \cosh^2\rho$
are automatic, and so the physical region is bounded only by
\be\label{BBvalid}
0 < \cosh^2\rho - a_1 \sin^2\theta - a_2 \cos^2\theta~. 
\ee
This is sufficient to ensure both of \eqref{eomrho3rep} and \eqref{eomtheta3rep} when $s_1s_3 = -1$. To summarize, the four-parameter family of giant string solutions that we call Type II is parameterized by four real numbers $(a_1,a_2,\rho,\theta)$ subject to \eqref{BBvalid} and $\rho \geq 0$ and $0 \leq \theta \leq \frac{\pi}{2}$.

The conserved quantities for this family of solutions follow from the general formulae 
(\ref{CFTJ}, \ref{CFTEP}) by inserting \eqref{BBbina}:
\bea\label{BBcharges}
e_R ~=~ q_R &=& \frac{(1-a_1)^2\sin^2\theta + (1-a_2)^2 \cos^2\theta + \sinh^2\rho}
{\cosh^2\rho - a_1 \sin^2\theta - a_2 \cos^2\theta}~, \\
e_L &=& \frac{\left((1\!-\!a_1)^2 \cosh^2\rho \!+\! a_1^2 \sinh^2\rho \right) \sin^2\theta
+ \left((1\!-\!a_2)^2 \cosh^2\rho \!+\! a_2^2 \sinh^2\rho \right) \cos^2\theta}
{\cosh^2\rho - a_1 \sin^2\theta - a_2 \cos^2\theta}~, \nn\\
q_L &=& \frac{\left( \!- (1\!-\!a_1)^2 \cosh^2\rho \!+\! a_1^2 \sinh^2\rho \right) \sin^2\theta
+ \left((1\!-\!a_2)^2 \cosh^2\rho \!-\! a_2^2 \sinh^2\rho \right) \cos^2\theta}
{\cosh^2\rho - a_1 \sin^2\theta - a_2 \cos^2\theta}~. \nn
\eea
Similarly to the Type I solutions, although the family of solutions has four parameters,
the charges are degenerate in that the family only covers a 3d subspace of the
4d charge space. Here the charge constraint defining the codimension one subspace is quite trivial:
\be\label{BBchargerel}
e_R = q_R~.
\ee
Thus Type II solutions are always $\frac14$-BPS, they have no physical region that breaks supersymmetry. 
The other BPS bound $e_L \geq |q_L|$ is built into the formulae
\eqref{BBcharges} because both $e_L \pm q_L$ are written as sums of squares.
For completeness, we note that the two $\mathbb{Z}_2$ symmetries that we have exploited to fix $s_1=s_2=+1$,
map this family to analogous families of Type II solutions with
$e_R = -q_R$, $e_L = q_L$ and $e_L = -q_L$, respectively.

The 3-to-4 map \eqref{BBcharges} can be inverted to write $\rho$ in terms of the
three independent charges:
\be\label{BBrho}
\cosh 2\rho = \frac{e_L-1}{e_R-1}
\quad \Leftrightarrow \quad
2\sinh^2\rho = \frac{e_L-e_R}{e_R-1}~.
\ee
This is the same as the Type I formula \eqref{eqn:rhoIEREL}, except for the interchange $L\leftrightarrow R$. 
Again, $\theta$ cannot be written in terms of charges, so it acts as the fourth parameter that can distinguish different solutions with the same value of charges.
Thus, the Type II giant string solutions can also be parametrized by four
\emph{physical} parameters, namely $e_R=q_R$, $e_L$, $q_L$ and $\theta$.

\begin{figure}
\begin{center}
\includegraphics[width=0.48\textwidth]{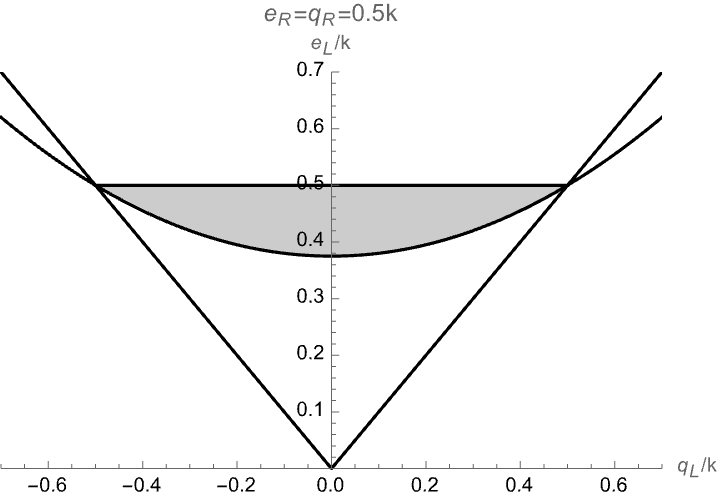}
\includegraphics[width=0.48\textwidth]{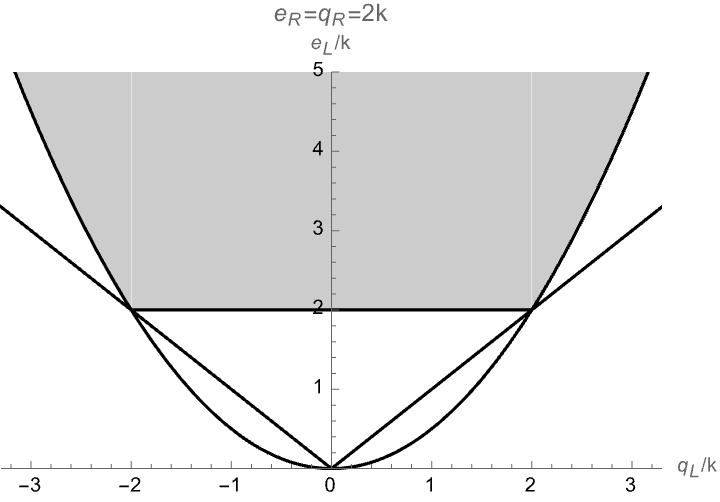}
\caption{\label{Fig:BBELJL} Allowed range of the charges $(q_L,e_L)$ for
Type II solutions. The two panels have $e_R=q_R = \frac12$ (left) and $e_R=q_R = 2$ (right). 
They qualitatively represent the cases $e_R=q_R < 1$ and $e_R=q_R > 1$.}
\end{center}
\end{figure}

From \eqref{BBrho} we infer that if $e_R > 1$ then $e_L \geq  e_R$, and vice versa.
Also, from \eqref{BBcharges} we derive the bound:
\be\label{BBchargeineq}
\frac{e_L}{2} - \left( \frac{q_L}{2} \right)^2 \geq \frac{e_R}{2} - \left( \frac{e_R}{2} \right)^2~.
\ee
Interestingly, Type I solutions of the previous subsection always saturate this bound,
according to \eqref{CCchargerel}.
Combining the two bounds, we find the physical region for the charges $e_L$ and $q_L$ with
given $e_R = q_R$:
\bea\label{BBELJLrange}
\frac{e_L}{2} - \left( \frac{q_L}{2} \right)^2 \geq \frac{e_R}{2} - \left( \frac{e_R}{2} \right)^2~,
\quad \text{and} \quad
\begin{cases}
e_L \leq e_R~, & \text{(if $e_R < 1$)} \\
e_L \geq e_R~. & \text{(if $e_R > 1$)}
\end{cases}
\eea
We plot examples of \eqref{BBELJLrange} in Figure \ref{Fig:BBELJL}. As representatives of two 
qualitatively distinct cases, we consider $e_R = q_R = \frac12 < 1$ and $e_R = q_R = 2 > 1$.
It is clear from the plots that \eqref{BBELJLrange} implies $e_L \geq |q_L|$.

\subsubsection{The Limit with Enhanced Supersymmetry}

Type II giant string solutions are always $\frac14$-BPS.
They constitute a larger class of supersymmetric giant strings than Type I solutions,
because both $\frac14$-BPS limits of the Type I solutions cover only
2d subspaces of charge space,
whereas Type II solutions cover a 3d subspace \eqref{BBcharges}. We now 
further impose the $\frac12$-BPS limit on the Type II solutions. This gives two constraints on 
parameter space and leaves a two parameter family of solutions that cover a 1d subspace of charge space.

Because $e_R = q_R$ is always satisfied, the $\frac12$-BPS limit of the Type II solutions is reached by further 
imposing $e_L = q_L$. Figure \ref{Fig:BBELJL} plots the allowed values of $(q_L,e_L)$ given $e_R=q_R$ and
illustrates how two conditions are imposed by the one equation $e_L = q_L$: it  is only satisfied at two specific points at the corners---not a line---in
the 2d plane of $(q_L,e_L)$. The same conclusion is reached from the fact that both $e_L \pm q_L$ given in \eqref{BBcharges} are sums of two squares. 

From \eqref{BBELJLrange} we see that $e_R=q_R$, $e_L = q_L$ is possible only if
\be\label{BBSUSY1}
e_L = e_R = q_L = q_R~.
\ee
This implies $\rho=0$ via \eqref{BBrho}, but $\theta$ remains as a continuous parameter.
Therefore, the $\frac12$-BPS limit is a two-parameter family, one of which is the charges \eqref{BBSUSY1}.
Thus, the Type II family is larger than the $\frac12$-BPS limit of the Type I solutions,
which also obey \eqref{BBSUSY1} but have fixed $\rho$ and $\theta$.

Recall that for any solution, two $\mathbb{Z}_2$ symmetries map to related solutions
with $L \leftrightarrow R$ and/or $q_{L,R} \leftrightarrow -q_{L,R}$. Thus, we have Type II giant string solutions with:
\be\label{BBSUSY2}
e_L = e_R = |q_L| = |q_R|~.
\ee
We see again that all $\frac12$-BPS solutions require $e_R = e_L$. This was also the conclusion from 
\eqref{CCSUSY3} which applies to both $\frac12$-BPS limits of Type I solutions.

\section{Thermodynamics of Giant Strings}
\label{sec:giantthermo}

The discussion of AdS$_3$ black hole thermodynamics in section \ref{sec:thermo} concluded that the dominant instability
is towards emission of particles that are chiral primaries on both left and right sides, i.e. $\frac12$-BPS particles.
However, as we have noted repeatedly, it may be too generous to assert the existence of $\frac12$-BPS particles
($e_R = q_R$ and $e_L = q_L$) with arbitrary values of $q_R$ and $q_L$.
For example, in the supersymmetric sigma model on $K3^k/S_k$, all $\frac12$-BPS particles
have $q_R \sim q_L$ in the large $k$ limit \cite{deBoer:1998kjm,deBoer:1998us}.

In this section, we study black hole stability with the assumption that, besides the AdS$_3$ black hole, only the giant strings discussed in the section \ref{sec:giantstrings} are relevant.
Because a giant string reduces the three-form flux at the black hole locus, it lowers the effective central charge, and this may have a significant cost in entropy.
In the analogous AdS$_5$ problem \cite{Choi:2024xnv,Choi:2025lck}, the black hole was allowed to be wrapped by
a single D3-brane that carries a macroscopic charge to minimize the cost.
We show that the giant strings we constructed in section \ref{sec:giantstrings} 
permit an analogous limit where a few giant strings carry charges that are so large $\sim k$ that they are 
comparable to those of the black hole.\footnote{We leave the study of interactions between the black hole and the giant strings to future work. Thus, we assume that the giant string solutions constructed in the global AdS$_3$ can coexist with the black hole, with the charges of the two components simply adding up.}

\subsection{Macroscopic Giant Strings}

Recall that the Type I giant strings carry charges \eqref{CCchargesabg} that are parametrized by
three parameters $\alpha$, $\beta$ and $\gamma$,
\bea\label{CCchargesabgrep}
\frac{e_R}{2} = \beta~, \quad
\frac{e_L}{2} = \alpha^2 + \beta (1-\beta) ~, \quad
\frac{q_R}{2} = \beta \gamma ~, \quad
\frac{q_L}{2} = \alpha~,
\eea
where the allowed values of the parameters are given by \eqref{condabg}:
\bea\label{condabgrep}
\begin{cases}
\beta < 0 & \text{: not allowed}~, \\
0 \leq \beta < \frac12 & \to |\alpha| \leq \beta~, \\
\frac12 < \beta & \to |\alpha| \geq \beta~,
\end{cases}
\qquad \text{and} \quad |\gamma| \leq 1~,
\eea
and visualized in Figure \ref{Fig:abgallowed}.
We can restrict to $q_L = 2\alpha \geq 0$ without loss of generality, 
because of the discrete $\mathbb{Z}_2\times\mathbb{Z}_2$ symmetry.

If charges $q_{R/L} \sim \sqrt{k} \gg 1$, the BPS inequality $e_{R/L} \geq |q_{R/L}|$ demands that 
the energies $e_{R/L}$ also scale at least as $\sqrt{k}$.
The converse is not necessarily true.
Therefore, it suffices for the study of macroscopic Type I giant strings to consider
those whose energies scale as $\sqrt{k}$.\footnote{In section \ref{sec:giantstrings},
charges of the giant strings were defined to include a factor $\sqrt{k}$, see \eqref{rescalesqrtk}.
We let the charges of the macroscopic giant strings to scale explicitly as $\sqrt{k}$
so the underlying quantum numbers scale as $\sim k$.}
First consider the case where $e_R = 2\beta \sim \sqrt{k}$.
The allowed range \eqref{condabgrep} imposes $\alpha \geq \beta$, so $\alpha$ must be at least that large.
In this limit, the formula \eqref{CCrho} for $\rho$ reduces to
\be
2 \sinh^2 \rho = \frac{\alpha^2 - \beta^2}{\beta}~,
\ee
so \eqref{CCchargesabgrep} gives: 
\bea
\frac{e_L}{2} &=& \beta + \alpha^2-\beta^2 = \beta \left( 1+ 2\sinh^2\rho \right) = \beta \cosh 2\rho~, \nn\\
\frac{q_L}{2} &=& \alpha = \sqrt{\beta^2 + 2 \beta \sinh^2\rho} = \beta \sqrt{1 + \frac{2 \sinh^2\rho}{\beta}}~.
\eea
Thus the four charges of the macroscopic giant become
\be\label{CCmacro1}
(e_R,e_L,q_R,q_L) = 2 \beta \cdot \left(1, \cosh 2\rho, \gamma , \sqrt{1+\frac{2\sinh^2\rho}{\beta}} \right)~.
\ee
The coordinate $\rho$ must be finite, so that $e_L$ of the giant string scales precisely as $\sim \sqrt{k}$, not as a larger power of $k$. That is required because the $e_L$ carried by the giant string is bounded from above by the energy of the entire system, which scales as $\sqrt{k}$ in the convention for the giant string charges.
As a result, the last component of \eqref{CCmacro1} becomes simply $q_L = 2\beta \cdot 1$.

In summary, we arrive at a class of macroscopic giant strings that is a subclass of the Type I solutions.
Their charges satisfy the following proportionality relations,
\be\label{CCmacro2}
(e_R,e_L,q_R,q_L) \propto \sqrt{k} \cdot \left(1, \cosh 2\rho, \gamma , 1 \right)~,~~
\qquad  |\gamma| \leq 1~~ \text{and}~~ \rho = {\cal O}(1)~.
\ee

Note that we started from $e_R \sim \sqrt{k}$, then $e_L \sim \sqrt{k}$ followed.
We now show that, by taking $e_L \sim \sqrt{k}$ first, we do not identify a more general class of macroscopic giants. 
Given large $e_L\sim \sqrt{k}$, $q_L$ can take any value as large as $e_L$, but \eqref{CCchargesabgrep} permits values as small as $\sqrt{e_L}$.
However, if $q_L \ll e_L$, the resulting giant string is uninteresting,
because then \eqref{CCchargerel} demands $e_R \ll e_L$,
so the giant string has charges proportional to $(e_R,e_L,q_R,q_L) \propto (0,1,0,0)$.
Strings with negligible momenta on the $S^3$ are never favorable for microcanonical considerations,
so we exclude them from our discussion.

Thus, if we assume $e_L \sim \sqrt{k}$, we must have $q_L\sim e_L \sim \sqrt{k}$. Then,
\be
1 \sim \frac{e_L-q_L}{q_L} = \frac{(\alpha-\beta)(-1+\alpha+\beta)}{\alpha} \approx
(\alpha-\beta) \cdot \frac{\alpha+\beta}{\alpha}~,
\ee
shows that $\alpha-\beta \sim 1$, while both $\alpha$ and $\beta$ scale as $\sqrt{k} \gg 1$.
It implies that $e_R = 2\beta$ scales as $\sqrt{k}$ as well, and from \eqref{CCrho} we again have
\be
2 \sinh^2 \rho = (\alpha - \beta) \cdot \frac{\alpha + \beta}{\beta} \sim 1~.
\ee
Thus, even if we take $e_L \sim \sqrt{k}$ first, it follow that $e_R \sim \sqrt{k}$ and $\rho$ is finite,
so we return to the class of macroscopic giant strings given in \eqref{CCmacro2}. Accordingly, this is the most general class of 
macroscopic giant strings that belong to the Type I solutions.

We now turn to the Type II solutions and first argue that $e_L \sim \sqrt{k}$ and $e_R \sim \sqrt{k}$ implies one another.
First, if $e_R = q_R \sim \sqrt{k} \gg 1$,
then the allowed values of $e_L$ are such that $e_L \geq e_R$, see right panel of Figure \ref{Fig:BBELJL}.
Again we do not allow $e_L$ to scale in higher powers of $k$ than $\sqrt{k}$, so this gives $e_L\sim\sqrt{k}$.
On the other hand, let we impose $e_L \sim \sqrt{k} \gg 1$ first.
According to the inequality \eqref{BBchargeineq}, unless $e_R = q_R$ also scales as $\sqrt{k}$,
$q_L$ must be small $q_L \ll e_L$.
In that case the four charges $(e_R,e_L,q_R,q_L)$ are proportional to $(0,1,0,0)$,
which we have argued is uninteresting.
So we have arrived at $e_R \sim \sqrt{k}$ starting from $e_L \sim \sqrt{k}$.
In conclusion, the only interesting macroscopic giant strings are those with
$e_L \sim \sqrt{k}$ \emph{and} $e_R \sim \sqrt{k}$ also for Type II solutions. 

Once we impose on the Type II solutions that both $e_L$ and $e_R$ scale as $\sqrt{k} \gg 1$ then 
\eqref{BBchargeineq} implies $|q_L| \leq e_R$.
Furthermore, from \eqref{BBrho},
\be\label{BBrhorep}
\cosh 2\rho \sim \frac{e_L}{e_R} \sim 1~.
\ee
Therefore, the four charges of the Type II macroscopic giant strings are described by
\be\label{BBmacro}
(e_R,e_L,q_R,q_L) ~~\propto~~ \sqrt{k} \cdot \left(1, \cosh 2\rho, 1 , \gamma \right)~,~~
\qquad |\gamma| \leq 1~~ \text{and}~~ \rho = {\cal O}(1)~,
\ee
in a manner that is very similar to the Type I subclass \eqref{CCmacro2}.

In summary, we have identified two classes of giant strings with macroscopic charges,
one from the Type I solutions \eqref{CCmacro2}, and one from the Type II solutions \eqref{BBmacro}.
By the discrete $\mathbb{Z}_2\times \mathbb{Z}_2$ symmetry discussed towards the end of section \ref{sec:symcharge},
there are analogous solutions whose $L$ and $R$ charges appear interchanged,
and additionally there are solutions where $q_{L,R}$ have the opposite signs.
Combining all, we conclude that there are macroscopic giant strings with charges that scale as:
\be\label{macrogiantall}
(e_R,e_L,q_R,q_L) ~~\propto~~
\begin{cases}
\sqrt{k} \cdot \left(1, \cosh 2\rho, \gamma, \pm 1 \right)~, \\
\sqrt{k} \cdot \left(1, \cosh 2\rho, \pm 1 , \gamma \right)~, \\
\sqrt{k} \cdot \left(\cosh 2\rho, 1, \gamma, \pm 1 \right)~,  \\
\sqrt{k} \cdot \left(\cosh 2\rho, 1, \pm 1 , \gamma \right)~,
\end{cases}
\ee
Here the parameter $\gamma\in [-1,1]$. The radial position of the giant string $\rho$ must be finite, 
it cannot be parametrically larger than $1$.

\subsection{Thermodynamics}

With the ingredients from the previous subsection, we now study the thermodynamics of a system composed of a black hole and a macroscopic giant string. Without loss of generality, we consider systems with both total charges $Q_R$ and $Q_L$ positive.
Then among the families \eqref{macrogiantall} of macroscopic giant strings, it is sufficient to consider:
\be\label{macrogiant}
(e_R, e_L, q_R, q_L) ~~\propto~~
\sqrt{k} \cdot\left(1, \cosh 2\rho_0, 1,1 \right) ~~\text{or}~~ \sqrt{k} \cdot\left( \cosh 2\rho_0,1, 1, 1 \right)~.
\ee
We take $\gamma=1$, since macroscopic strings with the largest charges possible are always the most favorable. In this special case, thermodynamics for Type I and Type II solutions becomes identical.
The radial position of the giant string is picked so the black hole background is sufficiently close to the global AdS$_3$ that the giant string solution applies.
We take $\rho_0$, a number of order $O(1)$, as the minimal value of such radial positions,
because greater $\rho$ always reduces $e_{R/L}$ of the black hole,
and one can see from the right half of Figure \ref{Fig:ERJRplane} 
that this always decreases the entropy of the system.
Concretely, the quantum numbers \eqref{macrogiant} correspond to chiral primaries on one side,
with $\rho_0$ parametrizing the departure from a chiral primary on the other side.  

To get oriented we consider the change in the left and right black hole entropy
\be\label{BHentropyLR}
S_{R} = 2 \pi \sqrt{\frac{k E_{R}}{2} - \frac{Q_{R}^2}{4}}~, \qquad
S_{L} = 2 \pi \sqrt{\frac{k E_{L}}{2} - \frac{Q_{L}^2}{4}}~,
\ee
independently, due to a small change of the form \eqref{macrogiant}. We could view 
this as emission of a small macroscopic giant string but, since we allow few of them, it is more appropriate 
to increase the proportionality constant in \eqref{macrogiant} by a small amount, for a giant string that is already in the system.
The change of the black hole charges is
\be
(\Delta E_{R}~,~\Delta E_L~,~\Delta Q_R~,~\Delta Q_{L}) =
-\lambda \left(1, \cosh 2\rho_0, 1,1 \right)~,
\ee
where $\frac{\lambda}{k}$ is small. The change in entropy, to the first order in $\frac{\lambda}{k}$, are
\bea\label{macroDeltaSRL}
\frac{\Delta S_{R}}{2\pi} &=& \lambda \cdot \frac{Q_R-k}{4\sqrt{\frac{kE_R}{2} - \frac{Q_R^2}{4}}}~, \nn\\
\frac{\Delta S_{L}}{2\pi}  &=& \lambda \cdot \frac{Q_L- k\cosh 2\rho_0}{4\sqrt{\frac{kE_L}{2} - \frac{Q_L^2}{4}}}~.
\eea
Therefore, for change of the black hole charges in the direction given by the first giant in \eqref{macrogiant}, the  black hole gains $S_R$ 
if $Q_R > k$ and gains $S_L$ if $Q_L > k\cosh 2\rho_0$. If both $S_R$ and $S_L$ increase, the black hole is clearly 
unstable towards a larger macroscopic giant string. Similarly, if they both decrease, it is unstable towards a smaller
giant string, provided that it already exists in the system. The most interesting is when the two contributions pull in opposite directions, 
yielding an equilibrium when: 
\be
\frac{\Delta S_{R}+\Delta S_{L} }{2\pi\lambda} = 
\frac{Q_R-k}{4\sqrt{\frac{kE_R}{2} - \frac{Q_R^2}{4}}} +
\frac{Q_L- k\cosh 2\rho_0}{4\sqrt{\frac{kE_L}{2} - \frac{Q_L^2}{4}}} = 0~.
\ee

This was all for the first of the giants in \eqref{macrogiant}, but it is straightforward to repeat the analysis for the second type: 
the subscripts $R$ and $L$ are simply exchanged. This gives the second equilibrium condition:
\be
\frac{Q_R- k\cosh 2\rho_0}{4\sqrt{\frac{kE_R}{2} - \frac{Q_R^2}{4}}} +
\frac{Q_L - k}{4\sqrt{\frac{kE_L}{2} - \frac{Q_L^2}{4}}} = 0~.
\ee
For both equations to be satisfied, we require
\be\label{macrogiantbalance}
\frac{Q_R+Q_L}{k} = 2\cosh^2\rho_0 ~, \qquad
E_R- \frac{Q_R^2}{2k} = E_L- \frac{Q_L^2}{2k}~.
\ee
The latter equation can be interpreted as the condition that $\beta_R = \beta_L$,
where $\beta_{R/L}$ are the potentials conjugate to $E_{R/L}$ defined by $\beta_{R/L} = \frac12 \beta (1 \mp \Omega)$.

Altogether, there is correlation between changes in $S_R$ and $S_L$, and the pattern is repeated from two types of giants. 
This complicates the analysis. The conditions \eqref{macrogiantbalance} defines a codimension-2 surface in the 4-dimensional charge space
on which changing the size of (or, emitting an extra piece of) either of the macroscopic giants \eqref{macrogiant} does not change the entropy.
Given the total charge of the system, there is a solution for the optimal size of both giants so that the remaining black hole charges belong to this equilibrium surface. Formally, this will be the most entropic configuration for the system composed of a black hole and the two species of macroscopic giants \eqref{macrogiant}. However, it may turn out that the solution found this way requires a negative number of giants, and for such a super-selection sector the equilibrium point is unphysical.

\begin{figure}
\begin{center}
\includegraphics[width=0.5\textwidth]{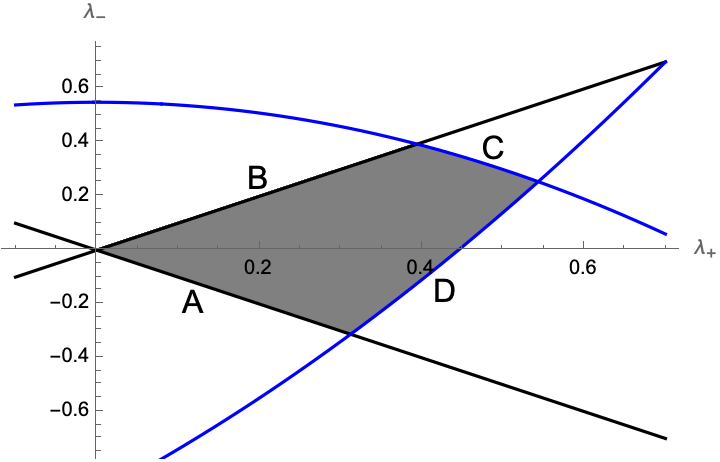}
\caption{\label{Fig:alpharegion} An example of allowed region for $(\lambda_+,\,\lambda_-)$
(in shade) plotted for $(E_R,\,E_L,\,Q_R,\,Q_L) = k(0.7,\, 1.4,\, 0.7,\, 1.5)$ and $\cosh^2\rho_0 = 1.5$.
Black lines depict the positivity bounds \eqref{alpha-bound1} and blue lines the
extremality bound \eqref{alpha-bound2}.
We label the four segments of the boundary as A, B, C and D.}
\end{center}
\end{figure}

We proceed more systematically, as follows. Denote by $\lambda_{R/L}$ the proportionality constant for the two macroscopic giants
\eqref{macrogiant} we account for, with $\lambda_R = 0$ or $\lambda_L=0$ meaning absence of the giant in question.
Given the total charges $(E_R,E_L,Q_R,Q_L)$ of the composite system, the black hole charges are:
\bea
E_{R,BH} &=& E_R - k\lambda_+ \cosh^2\rho_0 + k\lambda_- \sinh^2\rho_0~, \nn\\
E_{L,BH} &=& E_L - k\lambda_+\cosh^2\rho_0 - k\lambda_- \sinh^2\rho_0~, \nn\\
Q_{R,BH} &=& Q_R-k\lambda_+~, \nn\\
Q_{L,BH} &=& Q_L-k\lambda_+~,
\label{eqn:EQwalpha}
\eea
where we have defined $$\lambda_\pm = \lambda_R \pm \lambda_L~.$$
Factors of $k$ are products of $\sqrt{k}$ already present in \eqref{macrogiant}
and the implied scaling by another factor of $\sqrt{k}$ \eqref{rescalesqrtk} of the giants' charges.
Our goal is to maximize the black hole entropy:
\bea\label{BHentropyrep}
\frac{S_{BH}}{\pi} &=& \sqrt{2 k E_{R,BH} - Q_{R,BH}^2} + \sqrt{2 k E_{L,BH} - Q_{L,BH}^2}~,
\eea
subject to boundary conditions. First, the number (or size) of giant strings must be positive, $\lambda_{R/L} \geq 0$, so
\be\label{alpha-bound1}
-\lambda_+ \leq \lambda_- \leq \lambda_+~.
\ee
Furthermore, regularity of the core black hole imposes the extremality bounds \eqref{extbound} on its charges,
$E_{R/L,BH} - \frac{Q_{R/L,BH}^2}{2k} \geq 0$. This limits $(\lambda_+,\lambda_-)$ to:
\be\label{alpha-bound2}
-E_R + k\lambda_+ \cosh^2\rho_0 + \frac{(Q_R-k\lambda_+)^2}{2k}
~\leq~ k \lambda_- \sinh^2\rho_0 ~\leq~
E_L - k\lambda_+ \cosh^2\rho_0 - \frac{(Q_L-k\lambda_+)^2}{2k}~.
\ee
Plotting the four inequalities (\ref{alpha-bound1}-\ref{alpha-bound2}) on the  2d $(\lambda_+,\lambda_-)$-plane with typical 
parameters, we find the allowed region shown in Figure \ref{Fig:alpharegion}.

The maximization over the allowed region with boundaries can be solved via a standard procedure:
\begin{enumerate}[i)]
\item
Find extrema of the objective \eqref{BHentropyrep} in the 2d region within the boundaries.
\item
Confine the objective to each 1d segment of the boundary and find extrema therein.
\item
Evaluate the objective at each 0d intersection of the boundaries.
\end{enumerate}
Among the values obtained from i) through iii), the largest one is the global maximum.
As a result, the system may exhibit up to $9$ phases:
when the maximum is in the bulk of the region, on one of four segments of the boundary,
or on one of the four intersections.
However, only $4$ of the $9$ phases actually appear in this case.
The boundaries C and D represent where the core black hole is extremal on either side,
and that is when the corresponding piece of the objective \eqref{BHentropyrep} vanishes.
Therefore, the global maximum cannot arise on or at the ends of these boundaries, ruling out $5$ phases.

Given a super-selection sector with total conserved charges $(E_{R},\,E_{L},\,Q_{R},\,Q_{L})$,
we maximize the entropy \eqref{BHentropyrep} subject to the constraints
(\ref{alpha-bound1}-\ref{alpha-bound2}) following the procedure described in the last paragraph.
As a result, we identify four phases:
\begin{itemize}
\item \underline{Phase 0}:
when the maximum occurs at the intersection between boundaries A and B,
where $(\lambda_+,\lambda_-) = (0,0)$ (see Figure \ref{Fig:alpharegion}).
In this case, the system has the largest entropy when the entire charge is
carried by the black hole so there are no giant strings.
This indicates a stable black hole.

\item \underline{Phase L}:
when the maximum occurs on the boundary A, where $\lambda_+ = - \lambda_- \neq 0$,
or equivalently $\lambda_R = 0$ and $\lambda_L \neq 0$.
In this case, the system contains a macroscopic giant string with charges
$k \lambda_L (\cosh 2\rho_0,\,1,\,1,\,1)$ located at the radial coordinate $\rho_0$.
The black hole carries the remaining charges of the system.

\item \underline{Phase R}:
when the maximum occurs on the boundary B, where $\lambda_+ = \lambda_- \neq 0$,
or equivalently $\lambda_R \neq 0$ and $\lambda_L = 0$.
In this case, the system contains a macroscopic giant string with charges
$k \lambda_R (1,\,\cosh 2\rho_0,\,1,\,1)$ that is located at the radial coordinate $\rho_0$.
The black hole carries the remaining charges of the system.

\item \underline{Phase LR}:
the maximum occurs in the bulk of the allowed region, where $\lambda_R \neq 0$ and $\lambda_L \neq 0$.
In this case, two species of macroscopic giant strings coexist;
one with charges $k \lambda_L (\cosh 2\rho_0, 1, 1, 1)$,
and another with charges $k \lambda_R (1,\cosh 2\rho_0, 1, 1)$.
The black hole carries the remaining charges of the system.
\end{itemize}

We can find the boundary of black hole stability analytically, as follows.
First consider the boundary between Phases 0 and L.
The phase transition point is when the global maximum found on the boundary A,
where $\lambda_- = -\lambda_+$, approaches the point $\lambda_\pm = 0$
(see Figure \ref{Fig:alpharegion}).
Inserting \eqref{eqn:EQwalpha} with $\lambda_- = -\lambda_+$ in the black hole entropy
\eqref{BHentropyrep}, and then extremizing over $\lambda_+$:
\be
\frac{Q_R - k\lambda_+ - k \cosh 2\rho_0}
{\sqrt{2k(E_R - k\lambda_+ \cosh 2\rho_0) - (Q_R - k\lambda_+)^2}} +
\frac{Q_L - k\lambda_+ - k}{\sqrt{2k(E_L - k\lambda_+) - (Q_L - k\lambda_+)^2}} = 0~.
\ee
For the phase transition point, we set $\lambda_+ = 0$ in this equation and find: 
\be\label{bdy0vsLpre}
\frac{Q_R - k \cosh 2\rho_0}{\sqrt{2kE_R - Q_R^2}} +
\frac{Q_L - k}{\sqrt{2kE_L - Q_L ^2}} = 0~.
\ee
This condition can be written as an equation for $E_L$ as
\be\label{bdy0vsL}
2kE_L = Q_L^2 + \frac{(Q_L - k)^2}{(Q_R - k \cosh 2\rho_0)^2} \cdot (2kE_R-Q_R^2)~, \qquad
\left( Q_R - k \cosh 2\rho_0 \right) \left( Q_L - k \right) \leq 0~.
\ee
The latter inequality arises while squaring the equation and is obvious from \eqref{bdy0vsLpre}.
We have assumed that both extremality bounds $2kE_R - Q_R^2 \geq 0$ and $2kE_L - Q_L^2 \geq 0$ are met.
\eqref{bdy0vsL} defines the boundary between Phases 0 and L.
The analogous boundary between Phases 0 and R can be obtained easily, by exchanging $L \leftrightarrow R$:
\be\label{bdy0vsR}
2kE_L = Q_L^2 + \frac{(Q_L - k\cosh 2\rho_0)^2}{(Q_R - k)^2} \cdot (2kE_R-Q_R^2)~, \qquad
\left( Q_R - k \right) \left( Q_L - k\cosh 2\rho_0 \right) \leq 0~.
\ee
The point where the two boundaries \eqref{bdy0vsL} and \eqref{bdy0vsR} intersect
is where all 4 phases meet; it is the point of balance \eqref{macrogiantbalance}
where the black hole with the given charges is stable against any change in its charges
that correspond to both macroscopic giants.

\begin{figure}
\begin{center}
\includegraphics[width=0.44\textwidth]{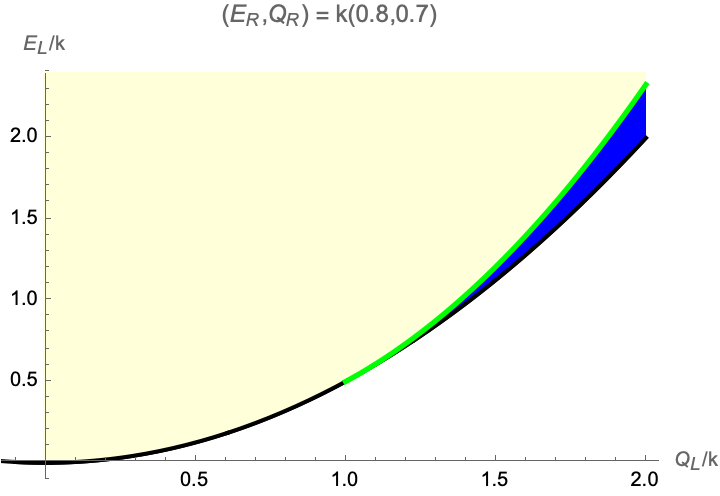}
\includegraphics[width=0.44\textwidth]{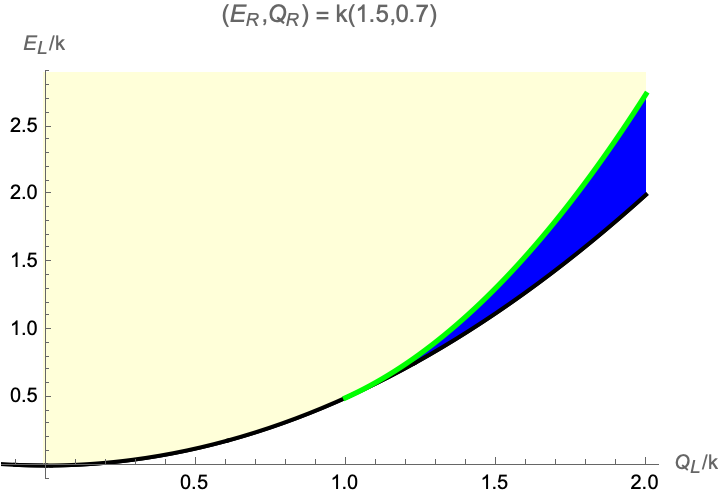}
\includegraphics[width=0.1\textwidth]{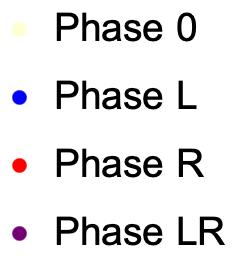}\\[10pt]

\includegraphics[width=0.44\textwidth]{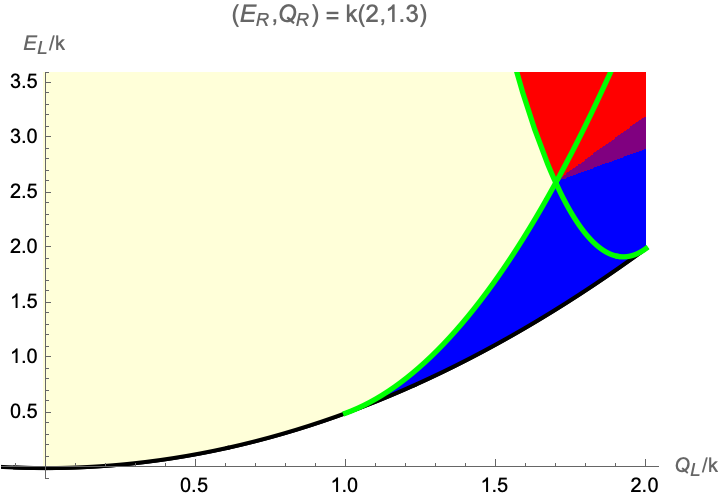}
\includegraphics[width=0.44\textwidth]{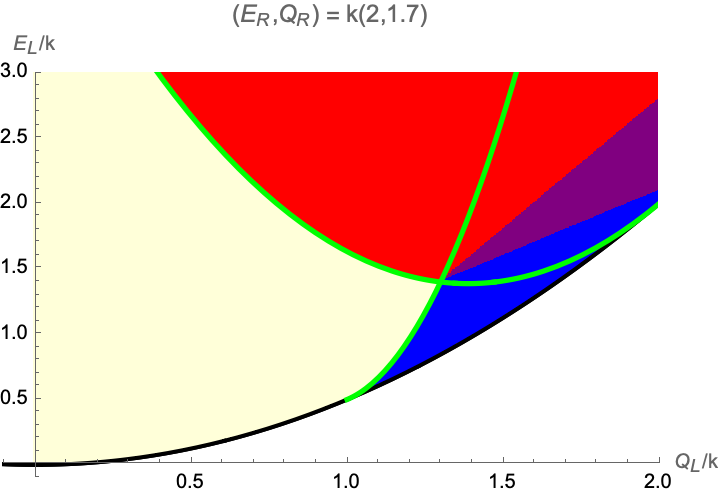}
\hspace{0.1\textwidth}

\caption{\label{Fig:macrophases} Phase diagrams on the $(E_L,\,Q_L)$-plane
for selected values of $(E_R,\,Q_R)$ and $\cosh^2\rho_0 = 1.5$.
Green curves show the boundaries between Phases 0 and L \eqref{bdy0vsL}
and between Phases 0 and R \eqref{bdy0vsR}.
The black parabola represents the black hole (left-)extremality bound \eqref{extbound}.
The four phases are colored in light yellow (Phase 0), blue (Phase L), red (Phase R) and purple (Phase LR).}
\end{center}
\end{figure}
In Figure \ref{Fig:macrophases}, we present phase diagrams on the $(E_L,\,Q_L)$-plane
for a selection of $(E_R,\,Q_R)$ and $\cosh^2\rho_0 = 1.5$. We plot the analytic boundaries \eqref{bdy0vsL} and \eqref{bdy0vsR} as green curves.
Note that the boundary \eqref{bdy0vsR} between Phases 0 and R is not present in the upper two panels,
because the inequality in \eqref{bdy0vsR} is not met for $Q_L < k\cosh 2\rho_0 = 2k$.
We make some general qualitative remarks based on the Figure:
\begin{enumerate}
\item Regardless of $(E_R,\,Q_R)$, smaller $Q_L$ makes it more likely that the system is in Phase 0
(the region in light yellow) where the black hole with given charges is stable.
This agrees with the basic principles highlighted below \eqref{macroDeltaSRL}.
Smaller $Q_L$ means that $S_L$ will not be gained by emitting either kind of giant.
A gain in $Q_R$ may overcome this, but smaller $Q_L$ makes it more demanding.

\item The analogous result with $L \leftrightarrow R$ is also illustrated by the Figures.
For smaller $Q_R$, the region for Phase 0 tends to be larger.
Note that panels in Figure \ref{Fig:macrophases} are placed in the order of (semi-)increasing $Q_R$.

\item On the right half ($Q_L > k$) of each panel.
In this part, the region just above the left-extremality bound
$E_L - \frac{Q_L^2}{2k} = 0$ (the black parabola), is always in phase L (in blue).
Thus it is favorable to emit the $\lambda_L$ macroscopic giant string (the second of \eqref{macrogiant}).
This can be understood from the entropy change due to its emission,
given by the $L \leftrightarrow R$ version of \eqref{macroDeltaSRL}.
The magnitude of $\Delta S_L$ blows up as the left-extremality bound is reached,
and the sign is positive if $Q_L > k$.

\item Conversely, well above the left-extremality bound,
the magnitude of $\Delta S_L$ is relatively suppressed,
so the behavior is predominantly determined by $S_R$.
It is precisely when $Q_R > k$ (the lower two panels) that $\Delta S_R$ in \eqref{macroDeltaSRL} is positive,
so it is favorable to emit a finite amount of the
$\lambda_R$ macroscopic giant string (the first of \eqref{macrogiant}),
placing the system in Phase R (in red).

\item The green curves represent the boundaries \eqref{bdy0vsL} and \eqref{bdy0vsR},
where it is marginally favorable to emit each type of macroscopic giant strings.
The boundary between Phase 0 and the other phases is part of these curves.
However, the other parts of the curves are entirely within the Phase L (blue) or R (red),
and do not coincide with the boundaries between Phases L and LR (purple), nor between R and LR.
So for example, there are systems that belong to the Phase L (it will only emit the $\lambda_L$ giant),
that lies on the right hand side of both green curves (emission of either giants would increase its entropy).
This is possible because, even if the system favors emitting both giants,
it may be the \emph{most} favorable to emit the $\lambda_L$ giant
all the way to the point where it is no longer favorable to emit any $\lambda_R$ giant.

\end{enumerate}

We finally comment on the discrete symmetries.
In this section we restricted the total charges of the system to $Q_R > 0$ and $Q_L > 0$.
Accordingly, we only considered the macroscopic giants \eqref{macrogiant} among
the bigger family of \eqref{macrogiantall} that have the maximal (positive)
values for both $q_R$ and $q_L$.
Using the discrete $\mathbb{Z}_2$ symmetries, one can repeat all the arguments for
any signs of the total charges $Q_R$ and $Q_L$ of the system.
For example, if $Q_R < 0$ and $Q_L < 0$, one should consider the macroscopic giants
with opposite $q_R$ and $q_L$ to \eqref{macrogiant}.
Then the entire thermodynamics that follow will be equivalent to what was presented in this section.

\section{Discussion}
\label{sec:discussion}

We conclude this article with a discussion of open problems that would be interesting for future research. We also take the opportunity to highlight
relations with other recent developments.

It has long been asked what states in CFT correspond to black holes in bulk AdS, and which do not. It is exciting that, in the last few years, a precise proposal has emerged \cite{Chang:2022mjp,Choi:2022caq,Choi:2023znd,Choi:2023vdm,Chang:2024zqi,Chang:2025rqy,Gadde:2025yoa}. The key technical tool is the cohomological classification of quantum states in SCFT. It gives a clear distinction between ``monotone" states, which exist in unchanged form also in the $N\to\infty$ limit, and ``fortuitous" states, that do not. The proposal is to interpret these classes of quantum states as supergravitons and black holes, respectively. 
Our work, and the development it belongs to, highlights the importance of additional CFT sectors
that describe core black holes ``dressed'' in various ways.
Hints of black holes dressed by gas of gravitons \cite{Choi:2023znd,Choi:2023vdm},
dressed by dual giant gravitons \cite{deMelloKoch:2024pcs},
and smooth horizonless solutions called ``singletons'' \cite{Hughes:2025car}
were identified from the fortuity, but there must be others.

We have stressed that the fate of black holes depends sensitively on the spectrum of the underlying theory, in a regime where it is not well understood. 
String theory in global AdS$_3$, offers a systematic theory that is well developed \cite{Giveon:1998ns,Maldacena:2000hw,Eberhardt:2018ouy}. In this approach short strings are confined, 
because of the AdS$_3$ ``box", while winding strings are scattering states that can make it to the conformal boundary of AdS$_3$. The latter may be important for the instability of black holes \cite{Martinec:2023plo}. Recent progress identified the deformation relating the spacetime CFT$_2$ of  AdS$_3$ string theory to the symmetric product orbifold CFT$_2$ \cite{Eberhardt:2021vsx,Chakraborty:2025nlb}. However, while this achieved a detailed description below the black hole threshold, the spectrum above remains unclear. 

To understand the spectrum better, we constructed macroscopic giant strings that seem novel, and may have interest in their own right. In higher dimensions, their analogues are central to the black hole phase diagram. 
However, more work is needed to understand the role of giants in AdS$_3\times S^3$. 

In this article, we take a somewhat low-brow approach, in an effort to find robust results that apply also in the absence of supersymmetry. Thermodynamics is certainly robust, and it has a dynamical interpretation via a standard application of the AdS/CFT correspondence. It is difficult to escape the conclusion that there are phase transitions at finite temperature, beyond those expected from wall-crossing and multi-center solutions. It would be interesting to characterize these phase transitions more precisely. 

\section*{Acknowledgements}

We thank
A. Belin,
S. Chakraborty, 
S. Choi,
O. Dias, 
R. Emparan, 
M. Guica,
C. Herzog, 
V. Krishna, 
K. Papadedimos
and K. Sharma
for discussions.
This work was supported in part by DoE grant DE-SC0007859 (FL),
a Rackham Predoctoral Fellowship and FWO project G094523N (SL).

\appendix

\section{Supersymmetric Giant Strings}\label{sec:reviewGS}

The goal of this appendix is twofold.
First, we recast the supersymmetry condition for the giant strings that derives from the
kappa symmetry analysis as holomorphicity of the string worldvolume in complex coordinates.
This is an adaptation to the AdS$_3 \times S^3$ giant strings of the connection \cite{Ashok:2008fa}
between the two formalisms of supersymmetry for the AdS$_5 \times S^5$ giant gravitons used
e.g. in \cite{Mandal:2006tk,Sinha:2006sh,Sinha:2007ni} 
and in \cite{Mikhailov:2000ya,Biswas:2006tj,Kim:2006he}.
Second, we use the holomorphicity of the worldvolume to present the simplest examples
of the supersymmetric giant string solutions.
They can be understood as lower-dimensional analogues of the
primitive giant graviton solutions in AdS$_5 \times S^5$ \cite{McGreevy:2000cw}
and dual giant gravitons \cite{Grisaru:2000zn,Hashimoto:2000zp}.
This appendix is mostly a review of existing literature such as \cite{Mandal:2007ug,Ashok:2008fa}.
We include it to present older results from a modern point of view, and to highlight the similarities between AdS$_3$ and AdS$_5$.

\subsection{Supersymmetry Conditions: General Analysis}

The condition for a D1-brane moving in a global AdS$_3 \times S^3 \times M_{\rm int}$ background
to preserve $4$ real ($2$ complex) supersymmetries was determined from kappa symmetry in \cite{Mandal:2007ug,Raju:2007uj}. 
The resulting supersymmetry condition was validated by showing that it is equivalent to solving
to the type-IIB Killing spinor equation.
We now show that it is further equivalent to the 2d worldvolume of the D1-brane being 
embedded in the 6d background AdS$_3 \times S^3$ \eqref{eqn:nearhAdS3}
by two functions that are both holomorphic and homogeneous.

The upshot of the kappa symmetry analysis in \cite{Mandal:2007ug} is that any 2d worldvolume in the AdS$_3 \times S^3$ background \eqref{eqn:nearhAdS3}
and \eqref{eqn:Bfields}, repeated here for convenience:
\bea\label{AdS3S3bginApp}
ds_6^2 &=& \ell^2 (d\rho^2 + \sinh^2\rho dy^2 - \cosh^2\rho dt^2)
+ \ell^2 (d\theta^2 + \sin^2\theta d\xi_1^2 + \cos^2\theta d\xi_2^2)~, \nn\\
C_2 &=& \ell T \left(\sinh^2 \rho \, dt \wedge dy
+ \left(\cos^2\theta + \frac{b-1}{2} \right) \, d\xi_1 \wedge d\xi_2 \right)~,
\eea
preserves 4 real supersymmetries if the vector
\be\label{Mandalnull}
{\bf n} = \partial_{t} + \partial_{y} + \partial_{\xi_1} + \partial_{\xi_2}~,
\ee
it tangent everywhere.

Meanwhile, one way to describe a 2d worldvolume in a 6d background is by imposing
4 real relations among 6 one-forms $dt, \cdots , d\xi_2$ at each point on the worldvolume,
so that all six differentials are parametrized by two independent worldvolume one-forms $d\tau$ and $d\sigma$.
It is useful for this appendix to define the vielbeins
\bea\label{3viel}
e^0 &=& - \cosh^2 \rho dt + \sinh^2 \rho dy~, \nn \\
e^1 &=& d\rho~, \nn \\
e^2 &=& \cosh \rho \sinh \rho (dy - dt)~, \nn \\
e^5 &=& d\theta~, \nn \\
e^6 &=& \cos \theta \sin \theta (d\xi_2 - d\xi_1)~, \nn \\
e^7 &=& \sin^2 \theta d\xi_1 + \cos^2 \theta d\xi_2~,
\eea
and mostly complex combinations thereof:
\bea\label{3mcc}
&& E^0 = e^0 + e^7~,\quad E^1 = e^1 - ie^2~,\quad E^5 = e^5 - ie^6~, \nn \\
&& E^{\bar{0}} = e^0 - e^7~,\quad E^{\bar{1}} = e^1 + ie^2~,\quad E^{\bar{5}} = e^5 + ie^6~, 
\eea
We refer to \emph{mostly} complex because $E^{\bar{1}}$ and $E^{\bar{5}}$ are complex conjugates of $E^1$ and $E^5$, but
$E^0$ and $E^{\bar{0}}$ are not complex conjugates of each other.
Contractions of the vector ${\bf n}$ with the one-forms $E^0, E^1, E^{\bar1}, E^5, E^{\bar5}$ all vanish, it is only the contraction with $E^{\bar0}$ that does not vanish.
Therefore, the supersymmetry condition that (\ref{Mandalnull}) is tangent to the worldvolume
is equivalent to the condition that the 4 real relations between the 6 one-forms
that define the worldvolume are between $E^0, E^1, E^{\bar1}, E^5, E^{\bar5}$,
but none of them involves $E^{\bar0}$.

We now connect the kappa symmetry condition to holomorphicity of the functions
that define the worldvolume. Suppose that
\bea
F(\rho, t, y, \theta, \xi_1, \xi_2) = 0~, \qquad 
G(\rho, t, y, \theta, \xi_1, \xi_2) = 0~,
\eea
on the worldvolume, for two complex functions $F$ and $G$ of the $6$ coordinates. Using subscripts to denote derivatives, we then have two complex differentials on 
the worldvolume: 
\bea\label{3dF}
dF &=& F_\rho d\rho + F_t dt + F_y dy
+ F_\theta d\theta + F_{\xi_1} d\xi_1 + F_{\xi_2} d\xi_2 \nn \\
&=& -\frac12 (F_t + F_y + F_{\xi_1} + F_{\xi_2}) E^{\bar0}
- \frac12 (F_t + F_y - F_{\xi_1} - F_{\xi_2}) E^{0} \nn \\
&& + \frac12 (F_\rho + i F_t \tanh \rho + i F_y \coth \rho ) E^1
+ \frac12 (F_\rho - i F_t \tanh \rho - i F_y \coth \rho ) E^{\bar1} \nn \\
&& + \frac12 (F_\theta - i F_{\xi_1} \cot \theta + i F_{\xi_2} \tan \theta ) E^5
+ \frac12 (F_\theta + i F_{\xi_1} \cot \theta - i F_{\xi_2} \tan \theta ) E^{\bar5} \nn \\
&=& 0~,
\eea
and similarly for $G$. More precisely, it is the pullback of the one-forms $dF$ and $dG$ onto the worldvolume that vanish.
The equations are equivalent to four real conditions, so they define a 2d worldvolume in the 6d spacetime.

The relation \eqref{3dF} simplifies in the complex coordinates defined by:
\bea\label{3Ccoord}
\Phi_0 = \cosh \rho \, e^{-it}~, ~~\Phi_1 = \sinh \rho \, e^{-iy}~,~~
Z_1 = \sin \theta \, e^{i \xi_1}~,~~ Z_2 = \cos \theta \, e^{i \xi_2}~.
\eea
The four complex coordinates are not independent, they obey two real relations:
\be
- |\Phi_0|^2 + |\Phi_1|^2 = -1~, \qquad |Z_1|^2 + |Z_2|^2 = 1~.
\ee
If the function $F$, when written as a function of the complex coordinates \eqref{3Ccoord},
is holomorphic in $(\Phi_0, \Phi_1, Z_1,Z_2)$,
then by chain rule from $\dd_{\bar\Phi_0} F = 0$ and so on, one can show that
\bea\label{3holocon1}
F_\rho - i \tanh \rho \, F_t  - i \coth \rho \, F_y &=& 0~, \nn \\
F_\theta + i \cot \theta \, F_{\xi_1} - i \tan \theta \, F_{\xi_2} &=& 0~.
\eea
If $F$ is further homogeneous in $(\Phi_0, \Phi_1, Z_1,Z_2)$ such that
the total degree in $\Phi_0$ and $\Phi_1$ is equal to the total degree in $Z_1$ and $Z_2$, then
\bea\label{3holocon2}
F_{t} + F_{y} + F_{\xi_1} + F_{\xi_2} &=& 0~.
\eea
When $F$ is holomorphic and homogeneous so that both \eqref{3holocon1} and
\eqref{3holocon2} are satisfied, then (\ref{3dF}) becomes a complex
linear relation between $E^0$, $E^1$ and $E^5$ only.
It also implies its complex conjugate, which is another linear relation between
$E^0$, $E^{\bar1}$ and $E^{\bar5}$.
Neither relation involves $E^{\bar0}$.

Therefore, if two complex functions $F$ and $G$ are
holomorphic in complex coordinates $(\Phi_0, \Phi_1, Z_1,Z_2)$
and are homogeneous with degree 0 in $(\Phi_0, \Phi_1, Z_1^{-1},Z_2^{-1})$,
$F=0$ and $G=0$ translate to two real relations between $E^0$, $E^1$ and $E^5$,
and another two real relations between $E^0$, $E^{\bar1}$ and $E^{\bar5}$.
Then for a 2d worldvolume in the 6d spacetime defined via $F=0$ and $G=0$,
the kappa symmetry condition of (\ref{Mandalnull}) being tangent to
the worldvolume everywhere, will be satisfied.
The D1-brane with such a worldvolume preserves 4 real supersymmetries.

We remark that holomorphicity and homogeneity of $F$ and $G$ is sufficient for supersymmetry of the worldvolume $F=G=0$, but not necessary.
However, the analogous construction of giant gravitons in AdS$_5 \times S^5$ \cite{Mikhailov:2000ya,Biswas:2006tj,Kim:2006he}
gives a class of supersymmetric D3-brane solutions that has proved very useful
in classifying and even quantizing the giant graviton solutions that preserve
$\frac18$ or more of the supersymmetries.

In the following, we present simple examples of supersymmetric giant string solutions in AdS$_3 \times S^3$. 
The giant wraps $S^1 \subset S^3$ and the dual giant wraps $S^1 \subset {\rm AdS}_3$.

\subsection{The Supersymmetric Dual Giant String}

Consider a D1-brane in the AdS$_3 \times S^3$ background \eqref{eqn:nearhAdS3}
whose worldvolume is defined by the two complex equations
\bea\label{3dgcon}
\Phi_0 Z_1 &=& \cosh \rho \sin \theta e^{i (\xi_1 - t)} = c_1~, \nn \\
 \Phi_0 Z_2 &=& \cosh \rho \cos \theta e^{i (\xi_2 - t)} = c_2~.
\eea
where $c_1$ and $c_2$ are complex numbers that must satisfy
$|c_1|^2 + |c_2|^2 \geq 1$.
Because $F = \Phi_0 Z_1$ and $G = \Phi_0 Z_2$ are holomorphic and homogeneous, 
the arguments in this appendix establish that this D1-brane preserves 4 real supersymmetries.
The geometry of the solution is summarized as follows:
\begin{itemize}
\item $\rho$ and $\theta$ are fixed to constants determined
$\cosh^2\rho = |c_1|^2+|c_2|^2$ and $\tan\theta = \frac{|c_1|}{|c_2|}$.

\item The other two $S^3$ coordinates $\xi_1$ and $\xi_2$ are linear functions of time $t$,
with offsets determined by the phases of $c_1$ and $c_2$.
Specifically, $\xi_{1,2} = t + {\rm arg}(c_{1,2})$.

\item $y$ is left unconstrained. In other words the brane extends along this direction.
\end{itemize}

One may take $t$ and $y$ as the worldvolume coordinates.
Then at each given time $t$, the azimuthal angles $\xi_{1,2}$ are both determined by
some linear functions of $t$.
So the giant string is pointlike in $S^3$, and this point is rotating with unit angular velocities
$\partial_t \xi_{1,2}=1$ at fixed $\theta$. In AdS$_3$, the string is at  fixed $\rho$ and the coordinate $y$ wraps a circle $S^1 \subset {\rm AdS}_3$ 
with radius $\sinh\rho$.

In the terminology of this paper, these supersymmetric dual-giant strings are special cases
of Type II solutions. Generally, the Type II solutions \eqref{BBbina} are parametrized by fixed values of $\rho$ and $\theta$,
and by $a_{1,2} = \dot\xi_{1,2}$, and then $b_{1,2}=\xi'_{1,2}$ are fixed via $b_{1,2} = 1-a_{1,2}$.
The supersymmetric dual-giants introduced here correspond to $a_1=a_2=1$ and $b_1=b_2=0$, with the
arbitrary $\rho$ and $\theta$ parametrized by $c_1$ and $c_2$ via 
$\cosh^2\rho = |c_1|^2+|c_2|^2$ and $\tan\theta = \frac{|c_1|}{|c_2|}$.

Using \eqref{BBcharges} we can compute the conserved quantities of the dual-giants:
\be\label{BBchargesdg}
e_R = q_R = e_L =1~, \qquad q_L = \sin^2\theta - \cos^2\theta~.
\ee
Note that $|q_L| \leq e_L = 1$.
Since $e_R = q_R$, it is clear that the dual-giants are supersymmetric,
as any Type II solution should be.
It is a codimension two subfamily of the Type II solutions defined by $e_R = q_R = 1$ and $e_L = 1$.
This subfamily is at the boundary of the space of Type II solutions in that the cases $e_R = q_R < 1$ and $e_R = q_R>1$
allow qualitative different physical region for $(q_L,e_L)$. The case $e_R = q_R = 1$ should be considered separately,
but it can be conveniently understood as limits from either side.
From both sides, i.e. in both panels of Figure \ref{Fig:BBELJL},
$e_R = q_R = e_L$ is realized along the horizontal line that bounds the allowed region.

\subsection{The Supersymmetric Giant String}

Consider a D1-brane in the AdS$_3 \times S^3$ background \eqref{eqn:nearhAdS3}
whose worldvolume is defined by the two complex equations
\bea\label{3gcon}
\Phi_0 Z_1 &=& \cosh \rho \sin \theta e^{i (\xi_1 - t)} = c_0~, \nn \\
\Phi_1 Z_1 &=& \sinh \rho \sin \theta e^{i (\xi_1 - y)} = c_1~,
\eea
where $c_0$ and $c_1$ are complex numbers that must satisfy
$|c_0|^2 - |c_1|^2 \leq 1$. The arguments of this appendix show that this D1-brane preserves 4 real supersymmetries
because $F = \Phi_0 Z_1$ and $G = \Phi_1 Z_1$ are holomorphic and homogeneous. The geometry of this solution is summarized as follows:
\begin{itemize}
\item $\rho$ and $\theta$ are fixed to constants determined by
$\tanh\rho = \frac{|c_1|}{|c_0|}$ and $\sin^2 \theta = |c_0|^2 - |c_1|^2$.

\item The other two AdS$_3$ coordinates $t$ and $y$ are linear functions of $\xi_1$
with offsets determined by the phases of $c_0$ and $c_1$ through $t = \xi_1 - {\rm arg}(c_0)$ and $y = \xi_1 - {\rm arg}(c_1)$.
Only one of $t$, $y$ and $\xi_1$ is independent so we can also express $y$ is a linear function of $t$. We can pick the origin of the coordinates so $t=y=\xi_1$. 

\item The azimuthal angle $\xi_2$ is unconstrained, so the brane extends along this direction.
\end{itemize}

One may take $t$ and $\xi_2$ as the worldvolume coordinates. Then, at each given time $t$, $y$ and $\xi_1$ are determined by linear functions of $t$.
So the giant string is pointlike in AdS$_3$, and the point is rotating with unit angular velocity
$\partial_t y=1$.
Meanwhile, it wraps the circle $S^1 \subset S^3$ parametrized by $\xi_2$,
whose radius is $\cos\theta$.
The radius is maximized when $\theta = 0$,
which corresponds to the maximal giant.
It is a distinct property of the giants, as opposed to dual-giants,
that the radius is bounded from above.

In the terminology of this paper, supersymmetric giant strings are also special cases of the Type II solutions, but the map is not obvious. 
The two independent worldvolume coordinates $t=y=\xi_1$ and $\xi_2$ must be obtained from the linear system \eqref{linearxi},
\bea\label{linearxirep}
\xi_1 = a_1 t + b_1 y~, \qquad
\xi_2 = a_2 t + b_2 y~,
\eea
where $t$ and $y$ are worldvolume coordinates. The coordinate change that is needed can be described as a singular limit: 
consider $- a_2 = b_2 - 1 = C$ and then take $C \to \infty$.
Then the second equation of \eqref{linearxirep} imposes $0=t-y$ for any finite value of $\xi_2$, while $\xi_2$ remains unconstrained.
In other words, we only look at a small $1/C$ portion of the worldvolume where
$t \approx y$ and $\xi_2$ is finite.
Then set $a_1 = b_1 = \frac12$. The first equation simply becomes $\xi_1 = t = y$,
as required for the description of the giant strings.
This choice of $(a_1,a_2,b_1,b_2)$ satisfies both conditions for the branches $\rho$B and $\theta$B:
\be
(a_1+b_1)^2 = (a_2+b_2)^2 = 1~, \qquad
(a_1-a_2)^2 = (b_1-b_2)^2 = (a_1b_2-a_2b_1)^2~.
\ee
To summarize, the giant strings are special cases of the Type II solutions
where we take a $1/C \to 0$ fraction of the string defined by
$a_1=b_1 = \frac12$ and $-a_2 = b_2-1 = C \to \infty$,
for arbitrary $\rho$ and $\theta$ determined by $c_0$ and $c_1$ via
$\tanh\rho = \frac{|c_1|}{|c_0|}$ and $\sin^2 \theta = |c_0|^2 - |c_1|^2$.
These parameters always satisfy the condition \eqref{BBvalid}, because of the large negative $a_2$.

Again, we can compute the conserved quantities of the giants from \eqref{BBcharges}.
Since the physical giant string corresponds to the $1/C \to 0$ fraction of the Type II solution
which is homogeneous, we divide the extensive quantities by $C$:
\be\label{BBchargesg}
\frac{e_R}{C} = \frac{q_R}{C} = \frac{q_L}{C} =1~, \qquad \frac{e_L}{C} = \cosh^2\rho + \sinh^2\rho~.
\ee
Note that $C = |q_L| \leq e_L$.
Since $e_R = q_R$, it is clear that the giants are supersymmetric, as any Type II solution should be.
It is the codimension two subfamily of the Type II solutions defined by $e_R = q_R = 1$ and $q_L = 1$.

\bibliographystyle{JHEP}
\bibliography{Instability.bib}

\end{document}